\documentclass[conference]{IEEEtran}
\IEEEoverridecommandlockouts
\usepackage{amsmath,amssymb,amsfonts}
\usepackage{algorithmic}
\usepackage{graphicx}
\usepackage{booktabs}
\usepackage{array}
\usepackage{textcomp}
\usepackage{xcolor}
\usepackage{url}
\usepackage{minted}
\usepackage{comment}
\usepackage[caption=false,font=footnotesize]{subfig} 
\usepackage{capt-of} 
\usepackage{enumitem}
\usepackage{float}
\usepackage[sorting=none]{biblatex}
\newif\ifsubmission
\submissiontrue

\ifsubmission
\newcommand{\mcnote}[1]{}
\else
\newcommand{\mcnote}[1]{\todo[color=violet!40,inline]{\textbf{MC:} #1}}
\fi

\def\BibTeX{{\rm B\kern-.05em{\sc i\kern-.025em b}\kern-.08em
    T\kern-.1667em\lower.7ex\hbox{E}\kern-.125emX}}
\begin{document}

\title{PICO: Performance Insights for\\ Collective Operations}

\newif\ifAuthors
\Authorstrue

\newcommand{\AFFSapienza}{Sapienza University of Rome, Rome, Italy}
\newcommand{\AFFKAUST}{KAUST, Thuwal, Saudi Arabia}
\newcommand{\AFFETH}{ETH Zurich, Zurich, Switzerland}

\newcommand{\AuthorBlock}{
    \author{
    \IEEEauthorblockN{
        Saverio Pasqualoni\IEEEauthorrefmark{1}$^{,}$\IEEEauthorrefmark{2},\hspace{0.3cm}
        Tommaso Bonato\IEEEauthorrefmark{3}, \hspace{0.3cm}
        Lorenzo Piarulli\IEEEauthorrefmark{1},
    }
    \IEEEauthorblockN{
        Torsten Hoefler\IEEEauthorrefmark{3},\hspace{0.3cm}
        Marco Canini\IEEEauthorrefmark{2},\hspace{0.3cm}
        Daniele De Sensi\IEEEauthorrefmark{1}
    }
        \IEEEauthorblockA{
            \small{
            \hfill
            \IEEEauthorrefmark{1}Sapienza University of Rome
            \hfill
            \IEEEauthorrefmark{2}KAUST
            \hfill
            \IEEEauthorrefmark{3}ETH Zurich
            \hfill
            }
        }
        
        \IEEEauthorblockA{\scriptsize{
            \texttt{\{saverio.pasqualoni,marco\}@kaust.edu.sa,}\;\;\texttt{\{piarulli,desensi\}@di.uniroma1.it,}\;\;\texttt{\{tommaso.bonato,torsten.hoefler\}@inf.ethz.ch}}
        }
    }
}

\newcommand{\AnonymousBlock}{
  \author{
    \IEEEauthorblockN{
      Anonymous Author(s)\\
      \{author(s)\}@institution.domain
    }
    \IEEEauthorblockA{
      \small{
        \hfill
        Anonymous Institution(s)
        \hfill
      }
    }
  }
}

\ifAuthors
  \AuthorBlock
\else
  \AnonymousBlock
\fi

\ifAuthors
\newcommand{\picourl}{\url{https://github.com/HLC-Lab/pico}}
\else
\newcommand{\picourl}{\url{https://github.com/xxxx/yyyy} omitted to comply with double blind rules}
\fi

\maketitle

\begin{abstract}
Collective operations are cornerstones of both HPC applications and large-scale AI training and inference, yet benchmarking them in a systematic and reproducible way remains difficult on modern systems due to the complexity of their hardware and software stacks. Existing suites primarily report end-to-end timings and offer limited support for controlled algorithm and configuration selection, fine-grained profiling, and capturing the runtime environment. We present PICO (Performance Insights for Collective Operations), an open-source framework that decouples portable experiment setup from platform execution, provides a backend-adaptive parameter selection interface across MPI and NCCL, supplies plain-MPI reference collective implementations, optionally instrumentable, and records the system configuration for reproducible comparisons. Evaluated on three major supercomputers, PICO shows that default collective algorithms and transport settings can be up to  $5\times$ slower than the best available choice. It provides diagnostic evidence by isolating topology sensitive algorithmic choices and, through instrumentation, reveals detailed algorithmic breakdowns. To assess end-to-end effects of benchmark-informed tuning and evaluate application-level impacts, we replay open-source LLM training traces in ATLAHS simulator with optimized collective profiles identified by PICO, achieving reductions in training times of up to $44\%$.
\end{abstract}


\section{Introduction} \label{sec:intro}
While recent High Performance Computing (HPC) systems have surpassed exascale performance~\cite{LLNL2025_ElCapitan, top500_2025_06}, the growing disparity between the available computational resources and data movement capabilities remains a critical obstacle. Processor performance continues to improve at a faster pace than memory and interconnect technologies, increasing the relative cost of communication. As HPC systems scale to ever-larger sizes, the efficiency of data transfers increasingly determines both application performance and overall system scalability~\cite{Liao2018, Exascale2014, Lu2022}.

Collective operations, being among the most communication-sensitive components of distributed-memory applications, are particularly affected by this trend and face increasing performance challenges on large-scale HPC clusters~\cite{hwang1998scalable}. Together with point-to-point communication, they form the backbone of traditional HPC workloads as well as AI training and inference tasks~\cite{Weingram2023}. At large scale, the cost of collective communication often dominates application runtime~\cite{laguna2019mpi_study, Weingram2023, balaji2009MPI_on_a_million}, making the design and optimization of efficient collective algorithms a key priority~\cite{desensi2024swing, sewell2024bruck, bienz2022bruckallgather, basu2024efficientalltoallcollectivecommunication, Sack2015collective, Dongkyun2025tidalmesh}.

Understanding and optimizing collective communication performance is difficult because collectives intertwine computation, memory movement, and network transfer, making it non-trivial to attribute bottlenecks to a specific subsystem~\cite{hunold2014reproducibleMPImicro, wu2024mscc}. This attribution problem is amplified by today’s heterogeneous environments: systems combine scale-up and scale-out fabrics with diverse topologies, multiple user-level communication libraries (e.g., MPI implementations and *CCL~\cite{nccl, rccl, accl, Weingram2023, gpugpuinterconnect}), and evolving network software stacks and APIs such as OFI and UCX/UCC~\cite{ucc, openucx-website, libfabric}. Moreover, performance is shaped by time-varying runtime conditions, such as congestion~\cite{zhang2020congestion}, load-balancing~\cite{bonato2025repsrecycledentropypacket} allocation policies~\cite{qureshi2020resource_allocation}, and task-to-node mappings~\cite{andrii2017taskmapping}, which introduce variance that is hard to predict, reproduce, or control.

Crucially, each collective operation can be implemented by multiple different algorithms, each one optimal for different combinations of node count and message sizes~\cite{thakur2005optimization, gpugpuinterconnect}: libraries often select among their available implementations with predetermined general heuristics. Moreover, performance gaps between different communication libraries, namely here GPU-aware MPI and *CCL libraries, can invert depending on runtime conditions, with one library outperforming the other depending on message size and node count~\cite{gpugpuinterconnect}. As a result, fair evaluation and reproducible characterization of collective algorithms require benchmarking methods that capture both system context and algorithm-level behavior.

Existing benchmarking tools such as OMB~\cite{OSU}, NCCL Tests~\cite{nccl_test}, Intel IMB~\cite{Intel_IMB_2021}, ReproMPI~\cite{hunsa_reprompi_2025}, and CommBench~\cite{hidayetoglu2024commbench} are effective for reporting end-to-end collective performance, but they only partially address the needs of modern systems. They generally lack fine-grained phase/step profiling and do not support straightforward, controlled comparisons of similar algorithms across different libraries. They also do not systematically record experimental conditions, such as node allocations, environment variables, software stack versions, and relevant hardware configuration, which complicates rigorous, reproducible evaluation.

For these reasons, we introduce \textbf{PICO (Performance Insights for Collective Operations)}, an open-source\footnote{\picourl}, modular, and extensible framework for benchmarking and diagnosing the performance of collective operations across multiple communication libraries. PICO contributes with (i) step-level instrumentation to attribute time to algorithmic phases, (ii) metadata-rich run capture to enable reproducibility and regression diagnosis, and (iii) backend-neutral reference collectives to isolate algorithmic differences from backend effects in a (iv) unified experiment specification spanning diverse backends (e.g., MPI and NCCL) for portable benchmarking.

We use PICO to analyze collective performance on different supercomputers: LUMI~\cite{Zwinger2023LUMI}, Leonardo~\cite{turisini2023leonardopaneuropeanpreexascalesupercomputer} and MareNostrum 5~\cite{banchelli2025introducingmarenostrum5europeanpreexascale}. We find that default collective algorithm selections can be $30$--$40\%$ slower than the best available alternative, and in the worst case deliver only $0.2\times$ of optimal performance. We leverage PICO’s comprehensive metadata capture to support a post-mortem analysis and identify root causes of diverging scaling behaviors of similar Broadcast algorithms, with the slower variant exhibiting a $2.5\times$ slowdown. Using PICO’s fine-grained instrumentation, we localize performance losses to different algorithmic phases, separating network-limited regimes from reduction and memory-movement limits.

Finally, to quantify end-to-end application impact, we replay open-source traces from LLaMA 7B and Mistral MoE using the ATLAHS simulator~\cite{bonato2025atlahs} with optimized collective profiles informed by PICO, achieving projected runtime reductions of up to $44\%$.

\section{Motivation and Requirements} \label{sec:requirements}
\begin{figure}[!t]
  \centering
  \includegraphics[width=\linewidth]{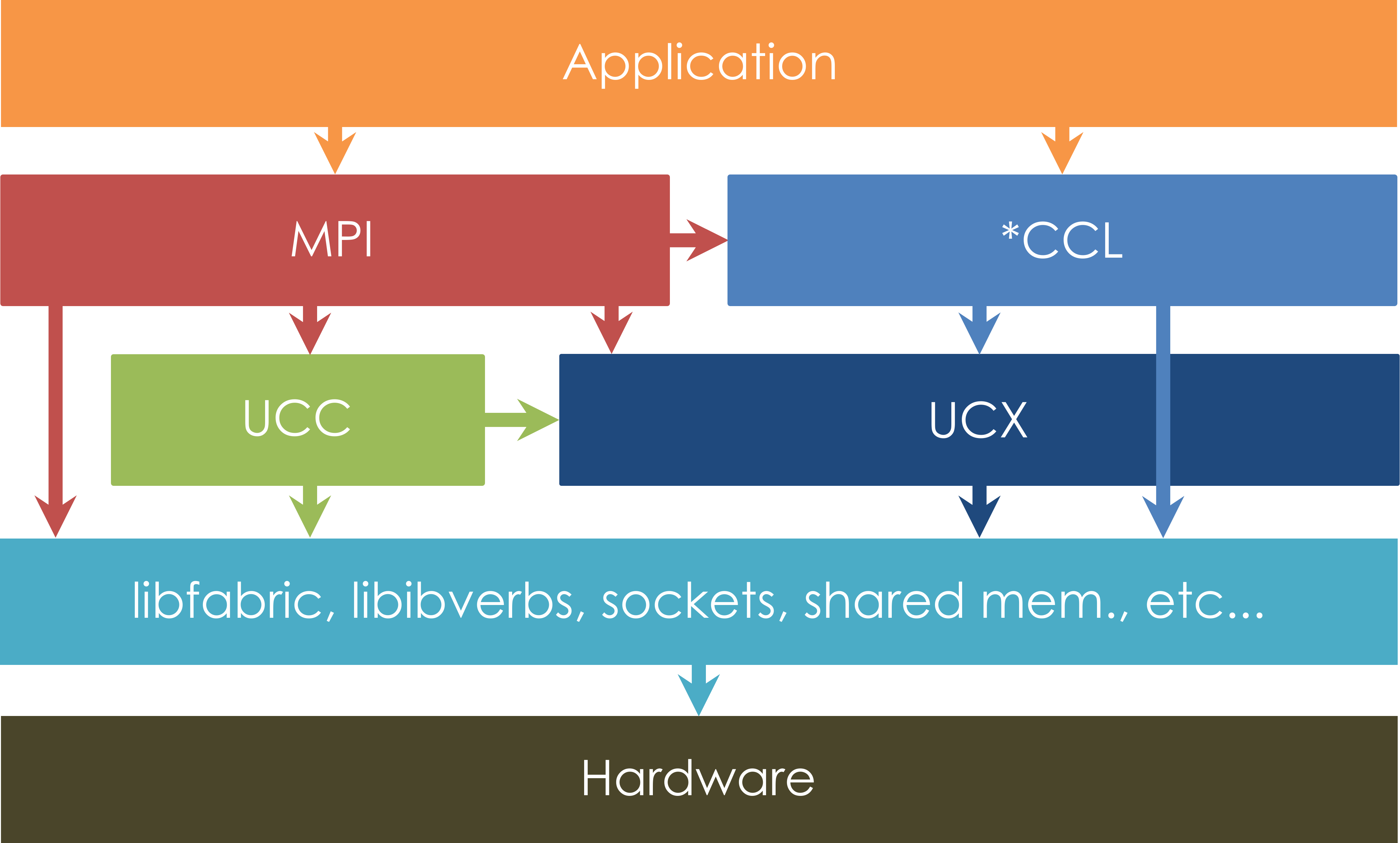}
  \caption{Simplified software stack hierarchy. Applications interface with high-level communication libraries (MPI, *CCL), while the high-level library typically relies on middleware layers, such as UCX or libfabric to access transport layer interfaces. Legacy versions of MPI and some *CCL libraries access directly the transport layer interfaces without relying on middleware APIs.}
  \label{fig:stack}
\end{figure}

Systematic, fair, and reproducible measurement of collective operations performance across modern HPC systems is increasingly difficult due to the tight coupling of the structured heterogeneity of hardware environments and the growing complexity of software stacks. The main challenges faced are:
\begin{description}[style=sameline,leftmargin=0em, itemsep=0.3em, font=\normalfont]
    \item[\textit{\textbf{C1}}]
    On the hardware side, modern platforms are typically built from multi-GPU nodes connected by high-bandwidth \textit{scale-up} fabrics (e.g., NVLink~\cite{NVIDIA_NVLink_Blog_2023}, InfinityFabric~\cite{Schor2018Zeppelin}, UALink~\cite{Synopsys2025UltraEthernetUALink}, UnifiedBuffer~\cite{zuo2025servinglargelanguagemodels}, or Scale-Up Ethernet (SUE)~\cite{Broadcom_SUE_Spec_2025}), whose bandwidth can exceed the \textit{scale-out} interconnect by up to an order of magnitude~\cite{gpugpuinterconnect}. At the same time, the size of scale-up domains is rapidly expanding (e.g., up to 72 GPUs in NVL72~\cite{NVIDIA_GB200_NVL72_2025} and even larger domains up to 384 GPUs~\cite{zuo2025servinglargelanguagemodels}). GPUs across nodes communicate over scale-out networks such as InfiniBand~\cite{Buyya_2002_infiniband}, Slingshot~\cite{De_Sensi_2020_slingshot}, or UltraEthernet~\cite{hoefler2025ultraethernetsdesignprinciples}, frequently deployed with tapered topologies (e.g., Dragonfly/Dragonfly+~\cite{kim_2008_dragonfly, Zwinger2023LUMI, frontier, De_Sensi_2020_slingshot, turisini2023leonardopaneuropeanpreexascalesupercomputer, shpiner_2017_dragonflyplus} or tapered fat-trees~\cite{banchelli2025introducingmarenostrum5europeanpreexascale, Juelich2025JUPITER_Tech, desensi2022noisecloudsinfluencenetwork}). Thus effective bandwidth and latency depends on whether communicating endpoints share the same scale-up domain and, for scale-out, whether they are within the same local switch domain or traverse oversubscribed global links~\cite{gpugpuinterconnect, De_Sensi_2020_slingshot}. Such non-uniform communication costs violate the homogeneous-link assumptions behind many traditional collective designs and motivate hierarchical and topology-aware collectives that explicitly exploit intra-node vs inter-node structure~\cite{bine, bienz2022bruckallgather, kandalla_2009_multi_ladder_allgather, larsson_2006_smp}.

    \item[\textit{\textbf{C2}}]
    On the software side, as illustrated in Fig.~\ref{fig:stack}, collective execution spans multiple layers (MPI implementations and *CCL libraries, as well as network stacks such as OFI/libfabric and UCX~\cite{libfabric, openucx-website}), each with tunable parameters and evolving behavior across versions. End-to-end timings include costs from network transfer, memory staging/movement, and reduction/computation, each dependent on one or more layer of the software stack. Moreover, results are sensitive to time-varying runtime conditions (e.g., congestion~\cite{zhang2020congestion}, allocation policies~\cite{qureshi2020resource_allocation}, and task-to-node mappings~\cite{andrii2017taskmapping}) as well as version/parameter changes across layers, which complicates reproducibility and regression diagnosis.

    \item[\textit{\textbf{C3}}]
    Beyond system heterogeneity, \emph{measurement methodology} itself is a source of systematic biases in collective benchmarking~\cite{hoefler-collmea}: precise process synchronization methods are required to ensure accurate results, but achieving this precision is a non-trivial task. A common approach is the use of barriers, but such constructs don't ensure that every single process enters the measured portion of the code at the same time: a process may exit the barrier before others, distorting measured runtimes~\cite{hunold2015impact}. In this regard the choice of barrier algorithm is important: some algorithms cause more skewing than others, with linear approaches (e.g. ring) being the worst due to their long propagation delay. An alternative approach to the use of barriers is represented by window-based schemes, where processes agree on a future start time. Those methods can reduce some barrier-induced artifacts, but they shift the problem to clock synchronization and drift~\cite{hoefler-collmea, hunold2015impact}.
\end{description}
\begin{figure}[!t]
  \centering
  \includegraphics[width=0.95\linewidth]{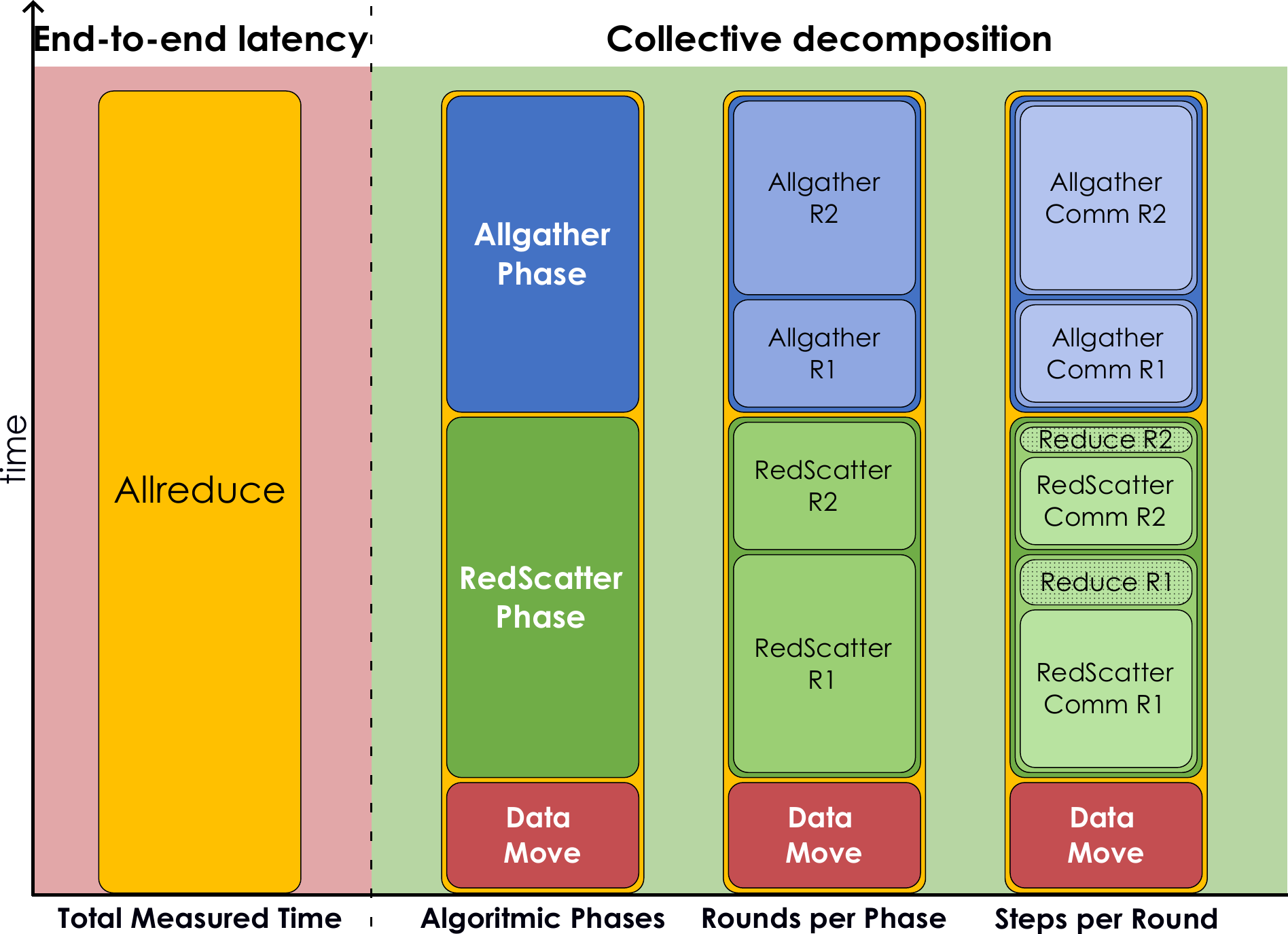}
  \caption{End-to-end benchmarks report a single latency for a collective (left), but execution decomposes into phases, rounds, and steps that mix communication, reduction/computation, and data movement (right). PICO targets this gap by enabling optional phase/step attribution and controlled baselines.}
  \label{fig:collective-gap}
\end{figure}
\subsection{State-of-the-Art Analysis}
Standard benchmark suites, including OSU-MicroBenchmark (OMB)~\cite{OSU}, Intel MPI Benchmark (IMB)~\cite{Intel_IMB_2021}, and NCCL Tests~\cite{nccl_test}, are effective at reporting end-to-end latency/bandwidth, but they are not designed for controlled, reproducible diagnosis. In particular, they provide limited support for (i) portable algorithm selection and cross-library baselining (often relying on manual environment-variable configuration), (ii) per step/phase measurements beyond aggregate timings, and (iii) lack mechanisms to record and store system's information and ensure reproducibility of the experiments.

This growing need for \emph{communication observability} is also reflected in industry efforts. NVIDIA recently introduced NCCL Inspector~\cite{nvidia2025ncclinspector}, a profiler-plugin that provides low-overhead, always-on, per-communicator and per-collective performance and metadata logging during real distributed AI workload runs, explicitly to help diagnose issues such as congestion and to correlate dips in compute throughput with collective performance. While valuable, such tooling is library-specific (NCCL-focused) and targets in-workload monitoring rather than portable, controlled benchmarking and backend-neutral baselining across stacks.

ReproMPI~\cite{hunsa_reprompi_2025} is a micro-benchmarking framework whose primary goal is correctness of measurements. It offers multiple synchronization methods and provides reference implementations of different collective algorithms but unfortunately it does not track the state of the system nor it aims to provide cross-stack compatibility and per step/phase measurements.

CommBench~\cite{hidayetoglu2024commbench} takes a significant step toward portability by introducing a library-agnostic API spanning MPI and *CCL collectives, but it does not target the same diagnostic and reproducibility objectives: it does not provide fine-grained observability of collective behaviours, metadata logging to track the state of the system is relatively minimal, and its usage model often requires writing low-level benchmarking logic, increasing manual effort for exploratory evaluation~\cite{hidayetoglu2024commbench}.

Netgauge~\cite{hoefler2007netgauge} introduces a modular benchmarking design that decouples benchmark “patterns” from communication “modules,” enabling extensible, comparable network/protocol measurements; however it is not designed for collective communications benchmarking.

Table~\ref{tab:soa-reqs-wide} summarizes this comparison, distinguishing capabilities that are natively supported as part of a tool’s integrated workflow ($\checkmark$) from those that are achievable only through external scripting, or ad hoc modifications ($\circleddash$) and those that are not supported at all ($\times$).
\begin{table}[!t]
\caption{Qualitative coverage of requirements (Sec.~\ref{sec:requirements:design}).\\
$\checkmark$: built-in; $\circleddash$: partial/manual; $\times$: not targeted.}
\label{tab:soa-reqs-wide}
\centering
\footnotesize{
\setlength{\tabcolsep}{2.2pt}
\begin{tabular}{lccccccc}
\toprule
 & \rotatebox{70}{OMB} & \rotatebox{70}{IMB} & \rotatebox{70}{NCCL-T/I} & \rotatebox{70}{CommBench} & \rotatebox{70}{NetGauge} & \rotatebox{70}{ReproMPI} & \rotatebox{70}{PICO} \\
\midrule
\textbf{R1} Fine grained profiling          & $\circleddash$ & $\times$       & $\checkmark$   & $\times$     & $\times$ & $\circleddash$       & $\checkmark$ \\
\textbf{R2} Backend-neutral references      & $\times$       & $\times$       & $\times$       & $\times$     & $\times$       & $\checkmark$   & $\checkmark$ \\
\textbf{R3} Portable spec \& control        & $\circleddash$ & $\circleddash$ & $\times$       & $\checkmark$ & $\checkmark$   & $\circleddash$ & $\checkmark$ \\
\textbf{R4} Automation \& amm. usability    & $\circleddash$ & $\circleddash$ & $\circleddash$ & $\checkmark$ & $\checkmark$   & $\checkmark$ & $\checkmark$ \\
\textbf{R5} Metadata-rich reproducibility   & $\times$       & $\times$       & $\times$       & $\times$     & $\circleddash$ & $\circleddash$ & $\checkmark$ \\
\textbf{R6} Extensibility across stacks     & $\circleddash$ & $\times$       & $\times$       & $\checkmark$ & $\circleddash$ & $\times$       & $\checkmark$ \\
\bottomrule
\end{tabular}
}
\end{table}
\subsection{Design requirements} \label{sec:requirements:design}
The central challenge in modern collective benchmarking is moving from end-to-end reporting to \emph{diagnosis}: controlled experiments that can explain \emph{why} a collective is underperforming and \emph{which} algorithmic step or subsystem dominates. Fig.~\ref{fig:collective-gap} illustrates this gap: many microbenchmarks typically report a single end-to-end latency, while the collective’s execution decomposes into phases, rounds, and steps (where a round is an iteration within a phase, and a step is a sub-operation within a round) that mix communication, reduction/computation, and data movement. Accordingly, our primary objective is \emph{fine-grained profiling} (\textit{R1}), enabled by \emph{backend-neutral, instrumentable baselines} (\textit{R2}) and a \emph{portable experiment specification} (\textit{R3}). The remaining requirements (\textit{R4--R6}) ensure reproducibility and practicality at scale.

\begin{description}[style=sameline,leftmargin=0em, itemsep=0.3em, font=\normalfont]
    \item[\textit{\textbf{R1}}]
    The framework must support fine-grained profiling at the granularity of algorithm phases, rounds and steps, enabling attribution of time to network transfer, memory movement/staging, and reduction/computation where applicable. Collective implementations must therefore be able to delineate regions of interest via explicit annotations that map measurements to semantically meaningful regions of the collective. Instrumentation must be \emph{optional} and impose no measurable overhead, within experimental noise, when disabled.

    \item[\textit{\textbf{R2}}]
    To enable controlled comparisons and meaningful attribution, the framework must include \emph{backend-neutral reference collectives} (e.g., plain-MPI implementations ported from major libraries) that isolate algorithmic differences from backend/transport effects (Sec.~\ref{sec:case-algodiff}) and can be instrumented at fine boundaries, providing consistent portable references for profiling without requiring modifications to implementation-specific internal code.

    \item[\textit{\textbf{R3}}]
    The framework must provide a portable, declarative specification for experiments (e.g., collective type, message sizes, scale, algorithm choice, and relevant backend parameters) so that the same experiment can be executed and compared across platforms and stacks with minimal platform-specific modifications. The specification must serve as a stable control interface that (i) selects among internal algorithm choices exposed by a given stack and (ii) exposes a set of relevant configuration parameters (Sec.~\ref{sec:case-tuning}), enabling controlled baselining across libraries without the need for specific per-experiment scripts.
    
    \item[\textit{\textbf{R4}}]
    The framework must support large-scale benchmarking campaigns while providing structured bookkeeping and post-processing. It must front-load complexity into an infrequent platform-setup step (e.g., module/environment configuration, backend discovery, scheduler integration), after which experiments can be executed from a portable specification. The system must automatically apply requested algorithm/knob settings, submit and manage jobs, collect results, and produce outputs in a standardized schema that support systematic comparison across runs, enabling recurring workflows such as tuning studies and regression checks.
    
    \item[\textit{\textbf{R5}}]
    Each run must capture sufficient metadata to reproduce and audit results and to diagnose regressions, including software stack versions (and build identifiers), selected backends/transports, relevant environment variables, hardware details (e.g., GPU/NIC model), and allocation/mapping context (e.g., node list and rank placement). Without this context, performance differences across runs, hardware or software updates are difficult to interpret. To keep large campaigns practical, the framework should support configurable metadata verbosity so that per-test metadata volume can be reduced when only minimal context is needed.
    
    \item[\textit{\textbf{R6}}]
    The framework must accommodate evolving communication stacks by providing a clear extension interface for adding new backends (MPI implementations or \emph{*CCL} libraries) and new collectives/algorithms: extensions to the framework should preserve a consistent end-to-end workflow. Moreover when a backend does not support a feature (e.g., algorithm selection or exposure to network layer parameters), the framework should degrade gracefully and still execute the experiment with a well-defined subset of functionality, without requiring all backends to implement the same sets of features.
\end{description}

\subsection{Workflows and Usability} \label{sec:requirements:human}
We selected these requirements to address the needs of three types of users. First, \emph{researchers} and \emph{collective algorithm developers} require backend-neutral baselines and fine-grained instrumentation capabilities (\textit{R1--R3}) to directly evaluate algorithmic differences and optimizations while ensuring reproducibility (\textit{R5}). Second, \emph{application developers and users} require automated exploration and structured result management to make tuning campaigns feasible and repeatable (\textit{R4}), together with sufficient metadata to interpret variability across allocations and software stacks (\textit{R5--R6}). Third, \emph{system administrators} require a repeatable workflow for regression testing across upgrades and configuration changes (\textit{R4--R5}).
\section{Architecture}
\label{sec:arch}
\begin{figure}[t!]
    \centering
    \includegraphics[width=\linewidth]{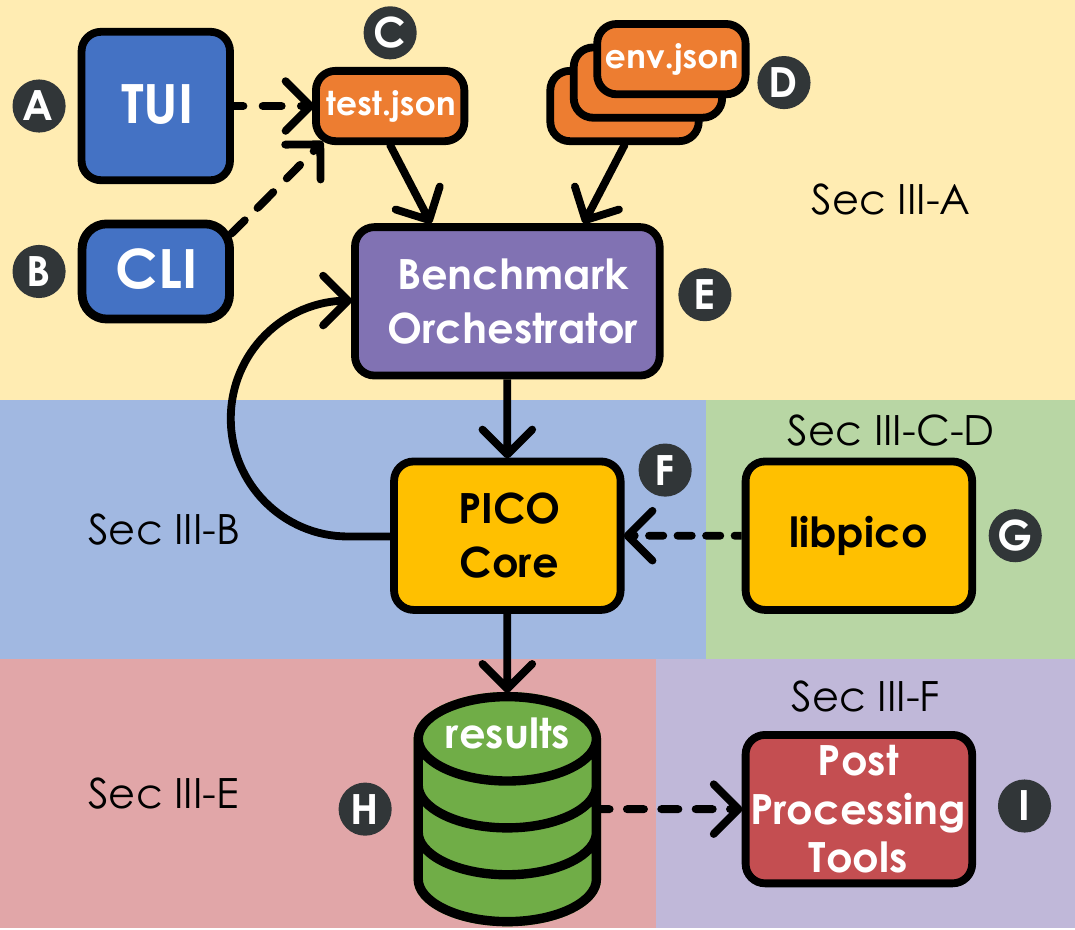}
    \caption{High-level PICO workflow. Users define experiments via a portable test descriptor (\texttt{test.json}) paired with a platform environment descriptor (\texttt{env.json}). The benchmark orchestrator resolves the descriptors, builds and launches \texttt{pico\_core} (and optional \texttt{libpico} baselines), and collects standardized results and metadata for post-processing and diagnosis.}
    \label{fig:pico-arch}
\end{figure}

PICO's core design principle is to decouple \emph{what to run} (a portable experiment description) from \emph{how to run it} on a given platform (reusable environment descriptors), while producing outputs that are both comparable and diagnosable. The resulting end-to-end workflow, as shown in Fig.~\ref{fig:pico-arch}, follows a simple pipeline: test and environment specifications are defined via descriptive files (\raisebox{-0.2em}{\includegraphics[height=1em]{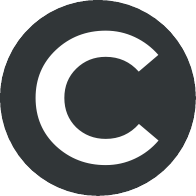}} and \raisebox{-0.2em}{\includegraphics[height=1em]{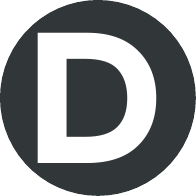}}), an orchestrator script (\raisebox{-0.2em}{\includegraphics[height=1em]{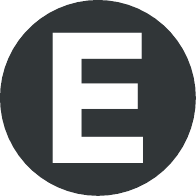}}) sets up the environment according to the test description and launches instances of a benchmarking core program (\raisebox{-0.2em}{\includegraphics[height=1em]{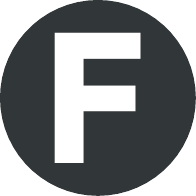}}) responsible for executing test instances and store performance data and metadata (\raisebox{-0.2em}{\includegraphics[height=1em]{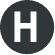}}) in a structured format. Post processing and visualization tools (\raisebox{-0.2em}{\includegraphics[height=1em]{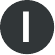}}) are used to analyze the data itself. The benchmarking core can optionally use a library (\raisebox{-0.2em}{\includegraphics[height=1em]{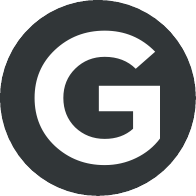}}) containing reference implementations  and instrumentation primitives.

\begin{figure*}[!t]
  \centering
  \includegraphics[width=0.495\linewidth]{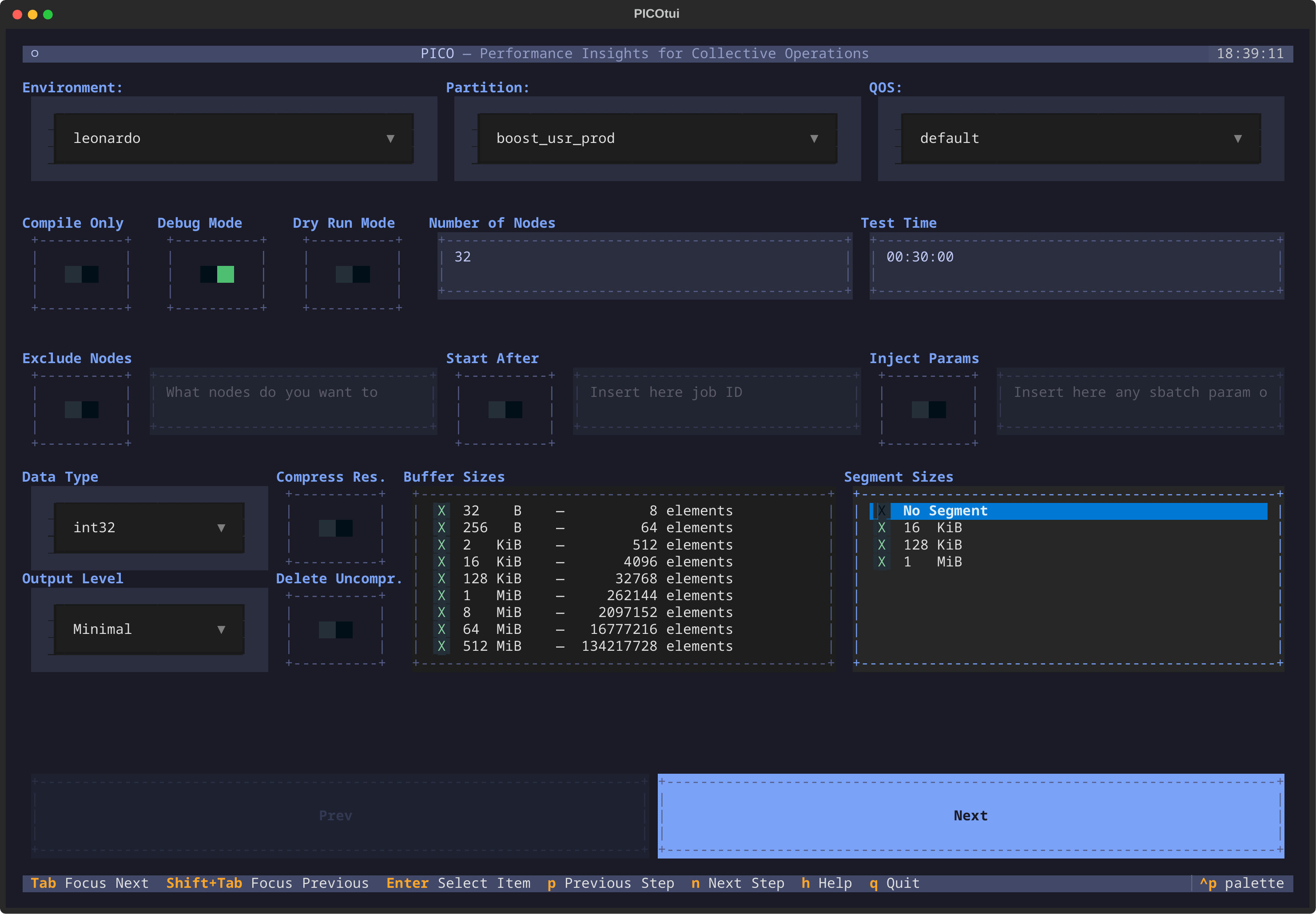}
  \includegraphics[width=0.495\linewidth]{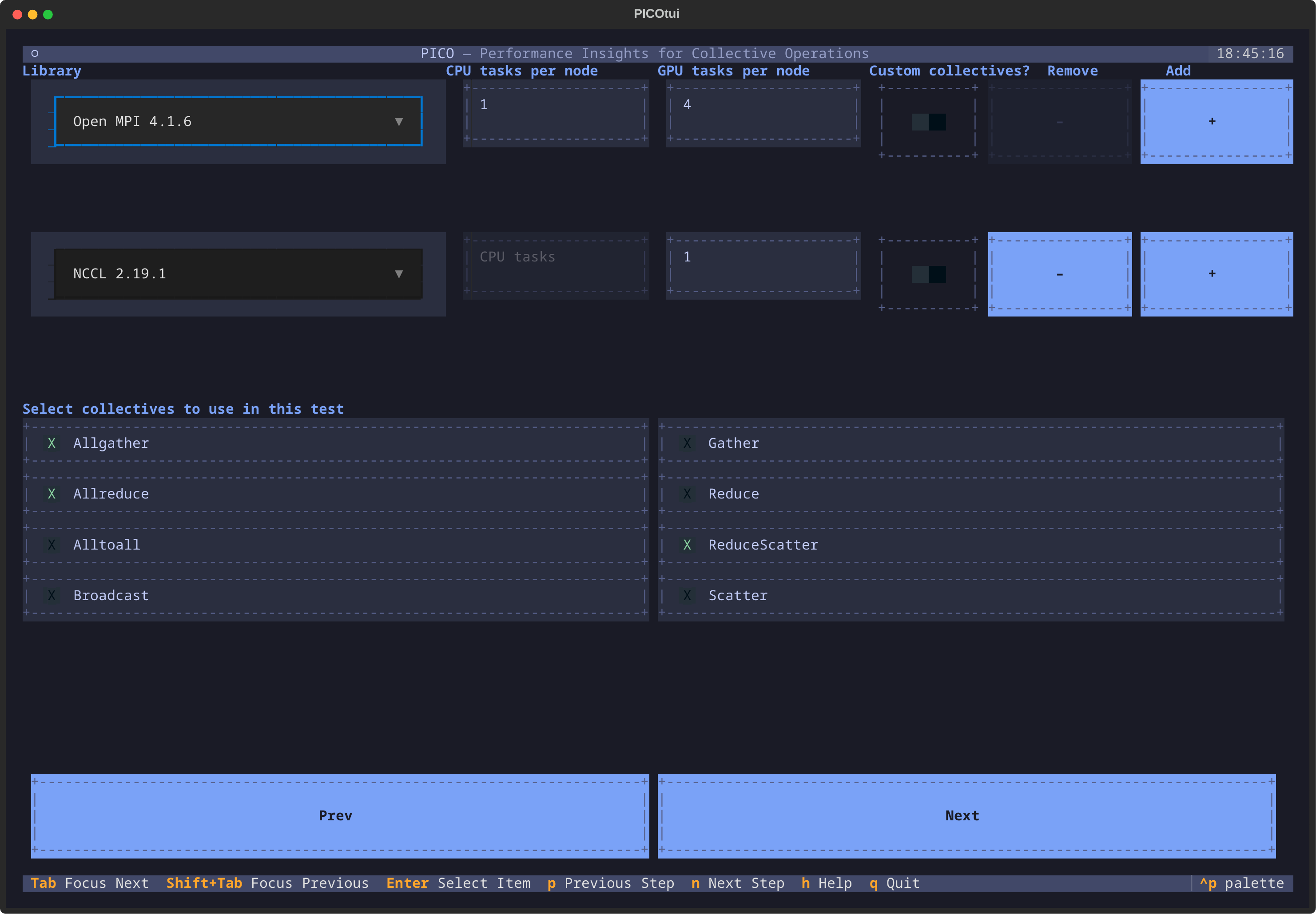}
  \caption{PICO terminal user interface (TUI) for interactive experiment specification. The TUI exposes available backends, collectives, algorithms, and supported control parameters for the current platform descriptor, and produces validated \texttt{test.json} files that can be executed by the orchestrator.}
  \label{fig:tui}
\end{figure*}

\subsection{Experiment specification and control plane} \label{sec:arch:spec}
PICO’s control plane provides a stable, portable interface for defining experiments and expressing backend control (algorithm choice and relevant parameters) through declarative descriptors. The framework translates this intent to backend-specific mechanisms without requiring per-experiment scripting. A platform descriptor (\texttt{env.json}; Fig.~\ref{fig:pico-arch}\raisebox{-0.2em}{\includegraphics[height=1em]{marker/d.pdf}}) records platform-specific capabilities and control mappings, while a portable test descriptor (\texttt{test.json}; Fig.~\ref{fig:pico-arch}\raisebox{-0.2em}{\includegraphics[height=1em]{marker/c.pdf}}) records experiment intent. Together, they realize \textit{R3} by making experiments executable and comparable across platforms with minimal intervention, and they enable \textit{R4} by allowing the orchestrator (Fig.~\ref{fig:pico-arch}\raisebox{-0.2em}{\includegraphics[height=1em]{marker/e.pdf}}) to execute large campaigns directly from descriptors.

To improve usability, set up configuration complexity is front-loaded into the infrequent creation of \texttt{env.json} descriptors. Those files define the local environment: available communication stacks (MPI implementations and selected \emph{*CCL} libraries), module/environment setup, scheduler and launcher templates (e.g., SLURM defaults), and backend-specific control mappings (e.g., which algorithm selectors and transport knobs are exposed and how to apply them). Given this platform context, users define a \texttt{test.json} that specifies the experiment in a backend-agnostic form (collective, message sizes, scale, and requested algorithm/parameter settings). Crucially, \texttt{test.json} does not encode cluster-dependent scripts; instead it encodes \emph{control intent} (e.g., ``use algorithm X'', ``set parameter Y'') that PICO resolves using \texttt{env.json}.

PICO provides both a terminal user interface (TUI; Fig.~\ref{fig:pico-arch}\raisebox{-0.2em}{\includegraphics[height=1em]{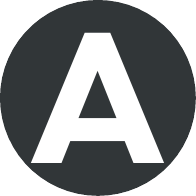}}) and a command-line interface (CLI; Fig.~\ref{fig:pico-arch}\raisebox{-0.2em}{\includegraphics[height=1em]{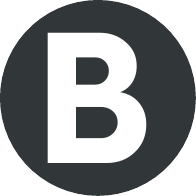}}) as front-ends to the same specification model. PICO's TUI, shown in Fig.~\ref{fig:tui}, is designed to guide the user into the discovery of available libraries and parameters: it presents what controls are available for the selected backend, applies defaults and validation, and outputs a self-contained \texttt{test.json} with the desired test configuration.

\subsection{Execution engine and backend adapters}
\label{sec:arch:exec}
PICO’s execution engine, PICO core (Fig.~\ref{fig:pico-arch}\raisebox{-0.2em}{\includegraphics[height=1em]{marker/f.pdf}}), runs on compute nodes inside the allocated job. PICO core is responsible for the timing-critical portion of benchmarking: initializing the communication context, applying requested controls when supported, executing the target collective(s) over the specified message sizes and scales, and emitting measurements and metadata in the standardized output format, using PICO’s internal barrier synchronization for timing alignment~\cite{hunold2014reproducibleMPImicro}.

PICO supports heterogeneous communication stacks (\textit{R6}) through a uniform backend adapter interface, with backend availability selected at compile time (e.g., \texttt{\#ifdef NCCL/CUDA} to support for GPU collectives). Each compiled-in adapter implements: (i) context initialization, (ii) mapping of abstract controls from \texttt{test.json} to backend-specific knobs when exposed, and (iii) collective execution and timing.

\subsection{Backend-neutral baselines}
\label{sec:arch:baselines}
\texttt{libpico} (Fig.~\ref{fig:pico-arch}\raisebox{-0.2em}{\includegraphics[height=1em]{marker/g.pdf}}) is a user-space library that provides reference implementations of collective algorithms. In the current version, \texttt{libpico} focuses on MPI: it provides plain-MPI implementations (built on point-to-point primitives and adapted from Open~MPI and MPICH) so algorithmic choices can be evaluated without, necessarily, relying on library internal collectives. This enables controlled, backend-independent comparisons of algorithmic behavior under identical experimental conditions. Sec.~\ref{sec:case-algodiff} will demonstrate the practical value of stable reference baseline to isolate algorithmic effects.

Importantly, \texttt{libpico} was designed to be extensible and allow developer to write and test directly new algorithms. It can be extended to support other communication libraries by implementing the corresponding backend signature and registering the implementation within \texttt{pico\_core}.
\begin{figure}[!h]
\begin{minted}[
  fontsize=\scriptsize,
  frame=lines,
  linenos,
  xleftmargin=1.2em,
  escapeinside=||   % <--- allows LaTeX between |...|
]{c}
int allreduce(const void *sbuf, void *rbuf, int count,
            MPI_Datatype dt, MPI_Op op, MPI_Comm comm) {
  PICO_TAG_BEGIN("init:mem-move");  |\label{ln:init-begin}|
  /* staging, temporary buffers allocations and copies */
  PICO_TAG_END("init:mem-move");    |\label{ln:init-end}|

  PICO_TAG_BEGIN("phase:redscat"); |\label{ln:rs-begin}|
  for (int step = 0; step < steps; step++) {
    PICO_TAG_BEGIN("redscat:comm", step); |\label{ln:rs-c-begin}|
    err = MPI_Sendrecv(...);
    PICO_TAG_END("redscat:comm", step); |\label{ln:rs-c-end}|

    PICO_TAG_BEGIN("redscat:reduction", step); |\label{ln:rs-r-begin}|
    err = MPI_Reduce_local(...);
    PICO_TAG_END("redscat:reduction", step); |\label{ln:rs-r-end}|
  }
  PICO_TAG_END("phase:redscat"); |\label{ln:rs-end}|

  PICO_TAG_BEGIN("phase:allgather"); |\label{ln:ag-begin}|
  for (int step = steps-1; step >= 0; step--) {
    PICO_TAG_BEGIN("allgather:comm", steps-1-step); |\label{ln:ag-c-begin}|
    err = MPI_Sendrecv(...);
    PICO_TAG_END("allgather:comm", steps-1-step); |\label{ln:ag-c-end}|
  }
  PICO_TAG_END("phase:allgather"); |\label{ln:ag-end}|
  return MPI_SUCCESS;
}
\end{minted}
\caption{Pseudo-code fragment illustrating tag-based fine-grained attribution in an instrumented Allreduce implementation: nested \texttt{PICO\_TAG\_BEGIN/END} markers annotate memory movement, algorithm phases, and per-step operations for fine-grained timing breakdowns.}
\label{lst:tag-fragment}
\end{figure}
\subsection{Tag-based instrumentation for fine-grained attribution}
\label{sec:arch:tags}
To move beyond aggregate end-to-end timings, PICO supports optional, tag-based instrumentation (\textit{R1}) for collectives implemented in \texttt{libpico} (including user-defined collectives). Tags delineate semantically meaningful regions of an implementation, such as data staging, algorithmic phases, and per-step communication/reduction, enabling fine-grained attribution of where time is spent. Because instrumentation lives in \texttt{libpico}, PICO provides this breakdown without modifying vendor communication stacks.

Instrumentation is expressed through lightweight macros (\texttt{PICO\_TAG\_BEGIN(...)} and \texttt{PICO\_TAG\_END(...)}), which can be used either flat or nested to capture hierarchical structure. The probes are optional, user-controlled, and inserted only at selected regions of interest. When enabled, PICO records timings for tagged regions and emits them using the same structured output model as other measurements; when disabled, the tag macros compile out to empty statements, preserving the behavior of standard benchmarking and leaving end-to-end tuning sweeps unaffected. The added cost per timing invocation was measured to be negligible (less than 100\,ns per tagged region). Fig.~\ref{lst:tag-fragment} illustrates the instrumentation of an Allreduce, in particular: (i) a memory initialization region (lines~\ref{ln:init-begin}--\ref{ln:init-end}), (ii) the two phase structure with Reduce-Scatter and Allgather of the algorithm (lines~\ref{ln:rs-begin}--\ref{ln:rs-end} and \ref{ln:ag-begin}--\ref{ln:ag-end}), and (iii) per-step regions inside each phase denoting communication and reductions (lines~\ref{ln:rs-c-begin}--\ref{ln:rs-c-end},  \ref{ln:rs-r-begin}--\ref{ln:rs-r-end}, and \ref{ln:ag-c-begin}--\ref{ln:ag-c-end}).

\begin{figure*}[!t]
  \centering
  \subfloat[Leonardo (Open MPI 4.1.6)]{\includegraphics[width=.315\textwidth]{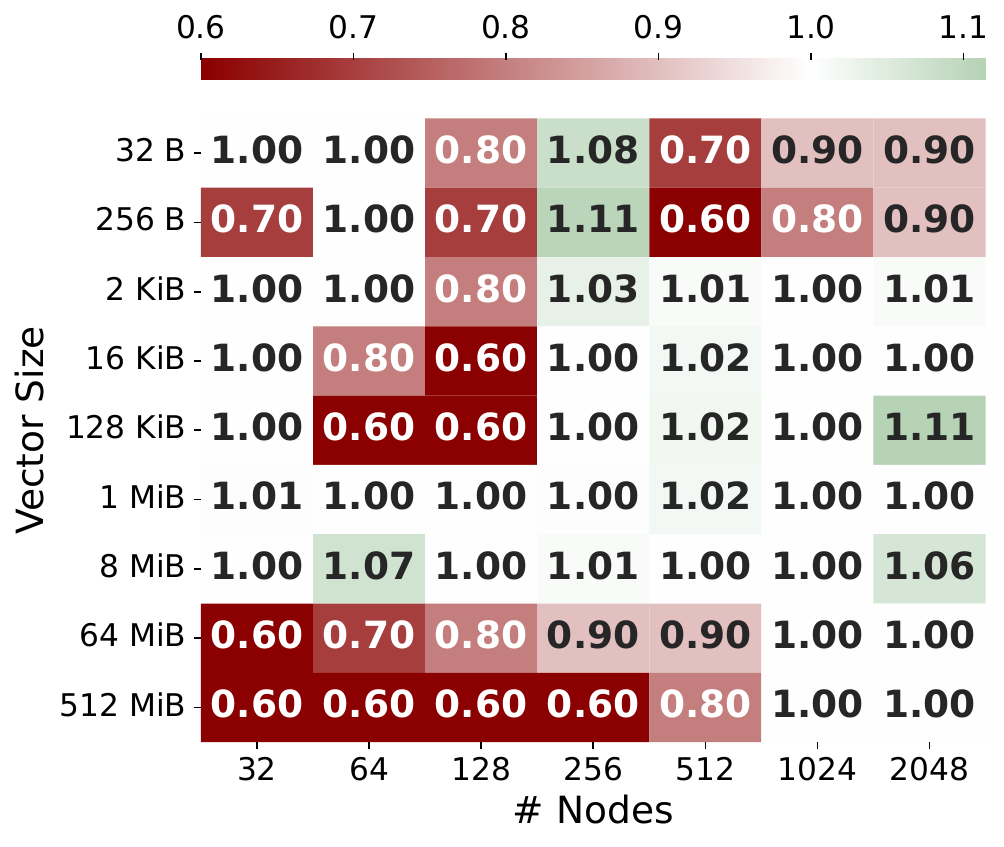}\label{fig:leonardo-default}}
  \subfloat[LUMI (Cray MPICH 8.1.29)]{\includegraphics[width=.315\textwidth]{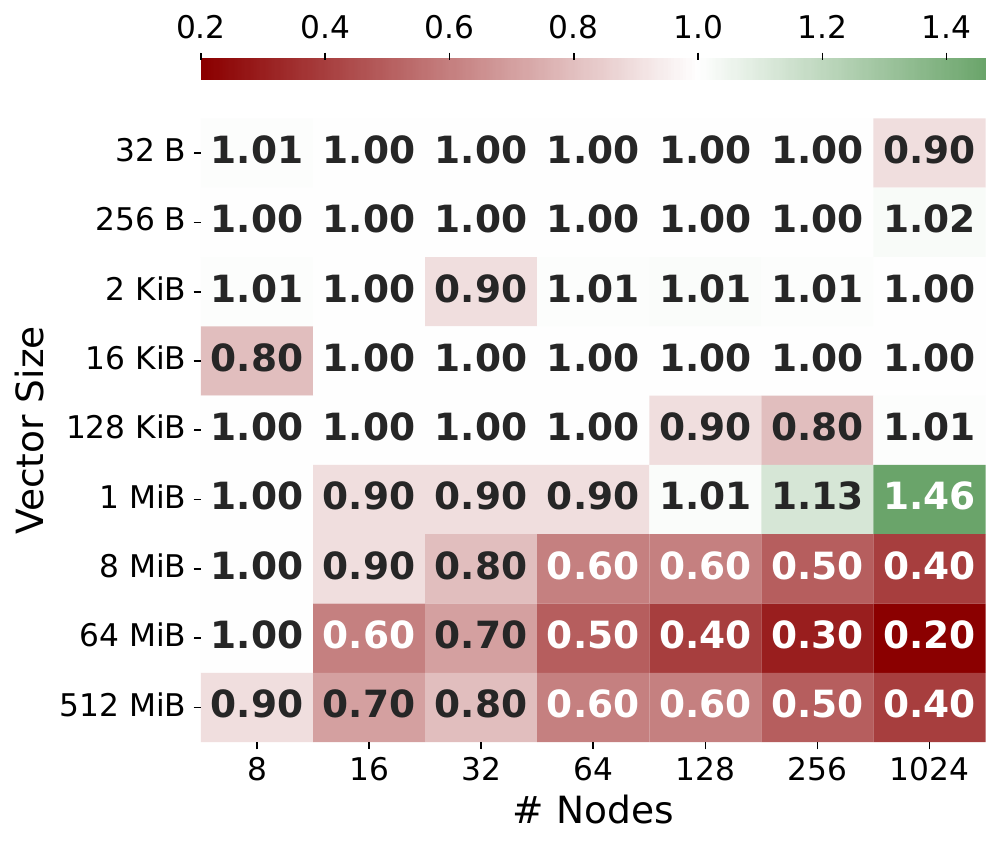}\label{fig:lumi-default}}
  \subfloat[MareNostrum 5 (Open MPI 4.1.5)]{\includegraphics[width=.315\textwidth]{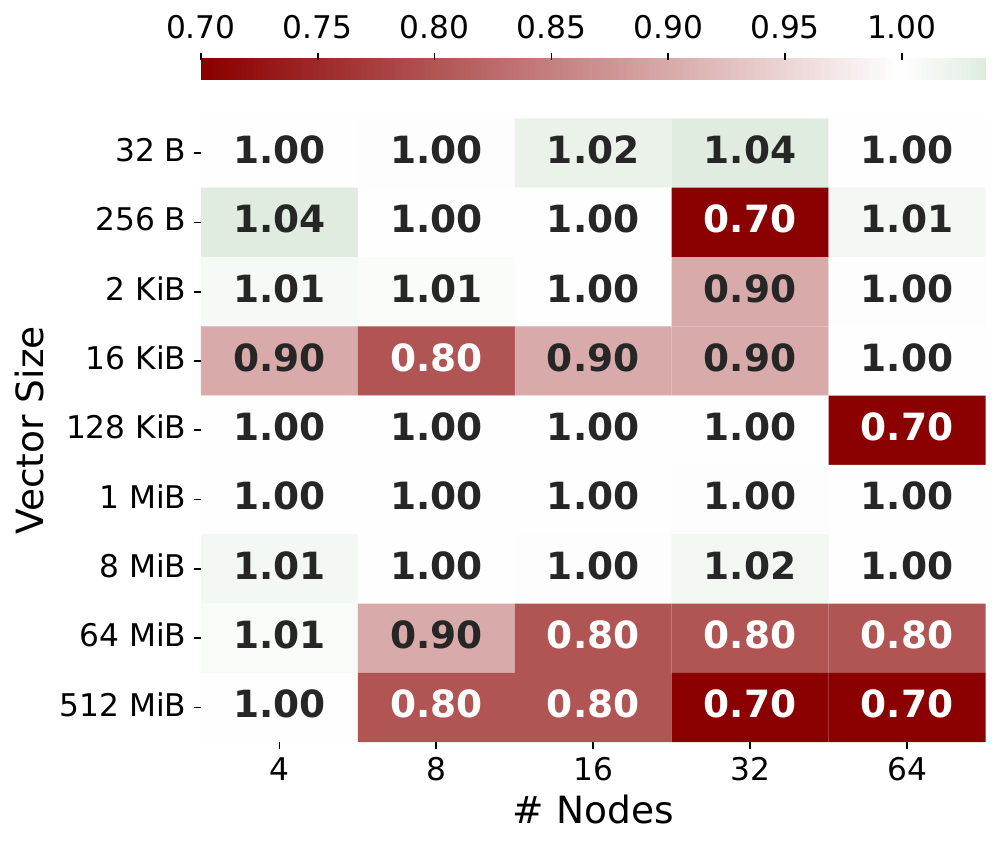}\label{fig:mn5-default}}
  \caption{Median best-to-default \emph{latency ratio} $r$ over the algorithm choices exposed by the communication library. In particular $r=\frac {t_{best}}{t_{def}}$ where $t_{def}$ is the algorithm automatically selected by the library and $t_{best}$ is the best performing algorithm not selected by default. Values $\tilde r<1$ indicate suboptimal choices.}
  \label{fig:default-choices}

\end{figure*}
\begin{table}[!t]
\caption{Result data granularity modes supported by PICO.}
\label{tab:result-modes}
\centering
\footnotesize{
\begin{tabular}{@{}lp{6.5cm}@{}}
\toprule
\textbf{Mode} & \textbf{Description} \\
\midrule
\texttt{Full}       & Stores all measurements for each rank and each iteration. \\
\texttt{Statistics} & For each iteration, stores aggregated statistics across ranks. \\
\texttt{Minimal}    & Records only the maximum value per iteration. \\
\texttt{Summary}    & Stores a single set of statistical aggregates over the iterations for the test point. \\
\texttt{None}       & Only \texttt{stdout} output with no values stored. \\
\bottomrule
\end{tabular}}
\end{table}

\subsection{Standardized results and metadata capture}
\label{sec:arch:results}
To satisfy \textit{R5}, PICO emits performance measurements and execution context in a standardized, human-readable output suitable for both large campaigns and post hoc diagnosis. Each campaign stores per-test measurements under a run directory, snapshots the resolved experiment specification (including the \emph{effective} control settings applied to the backend), and maintains a lightweight index to support automated traversal, aggregation, and comparison across runs.

Each \emph{test point} (collective type, message size, scale, backend, and control settings) is a separate record containing also timing data and identifiers; the schema is backend-agnostic and encodes both the \emph{requested} configuration (from \texttt{test.json}) and the \emph{effective} configuration after platform resolution (via \texttt{env.json}), preserving comparability even when controls are unsupported or mapped differently across stacks. To balance diagnostic depth and campaign scale, PICO supports configurable result granularity.

In addition, PICO records run context alongside performance data, including software stack versions/build identifiers, selected backends/transports, relevant environment variables and tuning knobs, hardware characteristics (e.g., GPU/NIC model), and allocation/mapping context (e.g., node list and rank placement). Metadata capture supports configurable verbosity so users can retain minimal context for broad sweeps while enabling richer capture for focused diagnostic runs.

\subsection{Analysis and diagnosis toolkit}
\label{sec:arch:analysis}
To help interpret collective performance on tapered, non-uniform interconnects, PICO provides a lightweight network traffic tracer that estimates how traffic is distributed across the cluster's different topology domains (e.g., Dragonfly groups). The tracer takes as input (i) allocation and rank-placement metadata captured per run (e.g., node list and rank mapping; \textit{R5}) and (ii) a topology description for the target system (e.g., node to switch-group membership and link hierarchy). It separates pairs of communicating ranks into categories based on their physical allocation (e.g., intra-node, intra-switch and inter-group) and returns an estimate of the utilization of network links by different collective algorithms. These estimates enable users to correlate observed performance with expected link congestion when comparing algorithms that trade local aggregation for reduced global traffic, or when diagnosing performance shifts caused by placement changes or topology-aware transport settings. It provides a topology-level estimate only, not a packet-accurate simulation of congestion, adaptive routing, or protocol behavior.

For convenience and amortized usability (\textit{R4}), PICO provides scripts that generate standard plots directly from the result schema, including heatmaps (e.g., message size vs.\ scale), line plots, and box/bar summaries across algorithms or backends. Because plots are derived from the same indexed schema used for campaign execution, visualization remains consistent across runs and can be integrated into automated tuning and regression pipelines.

\section{Evaluation}
\label{sec:eval}
This section evaluates PICO through a set of case studies designed to validate the requirements in Section~\ref{sec:requirements} and to demonstrate practical utility beyond end-to-end benchmarking. We organize the evaluation around four questions.
\begin{enumerate}
    \item How often does the library’s algorithm selection deviate from the best-performing choice for a given message size and process topology, and can PICO orchestrate controlled tuning campaigns? (Sec.~\ref{sec:case-tuning})
    \item Can \texttt{libpico} reference implementations isolate algorithmic effects and explain cross-platform differences? (Sec.~\ref{sec:case-algodiff})
    \item Can a more detailed instrumentation of a collective algorithm reveal actionable bottlenecks that are opaque under aggregate timings? (Sec.~\ref{sec:case-instr})
    \item Do benchmark-informed choices translate into application-level implications under realistic communication traces? (Sec.~\ref{sec:case-sim})
\end{enumerate}
Each case study shows what PICO makes possible \textit{because} it was designed around the design goals described in Sec.~\ref{sec:requirements:design}.

\subsection{Collective Tuning}
\label{sec:case-tuning}
Using PICO we conducted systematic sweeps of \texttt{MPI\_Allreduce} across three major European supercomputers, Leonardo (Open MPI 4.1.6), LUMI (Cray MPICH 8.1.29), and MareNostrum 5 (Open MPI 4.1.5), varying only the collective algorithm choice exposed by each communication stack while keeping all other settings fixed. Fig.~\ref{fig:default-choices} summarizes the outcome of these sweeps using the \emph{best-to-default latency ratio} $r = \frac{t_{\text{best}}}{t_{\text{def}}}$, where $t_{\text{def}}$ is the median runtime obtained under the backend default algorithm and $t_{\text{best}}$ is the minimum median runtime among the \emph{non-default} algorithms exposed by the backend for the same test point (message size, scale, and stack). Thus, $r<1$ indicates that the default choice is suboptimal (a faster non-default alternative exists), while $r>1$ indicates that the default is best among the exposed choices. Across all three systems, the heatmaps exhibit structured regions, often at larger scales and for specific message sizes, where defaults fall short of the best alternative by roughly $30$--$40\%$, and in the most pronounced case achieving only $20\%$ of the optimal performance.

Default selection heuristics are typically engineered to be conservative and broadly portable; as a result, they may fail to capture the platform-specific characteristics that matter in practice. This motivates systematic tuning of collective communication algorithms and their selection parameters. PICO's outputs can be used as empirical basis to tune libraries algorithmic choices: Open MPI supports overriding algorithm selection using \texttt{coll\_tuned} dynamic decision files~\cite{OpenMPI_CollTuned_5.0}, and analogous mechanisms exist across other MPI implementations and \emph{*CCL} libraries (via configuration files or environment variables).

\begin{figure}[!h]
  \includegraphics[width=\columnwidth]{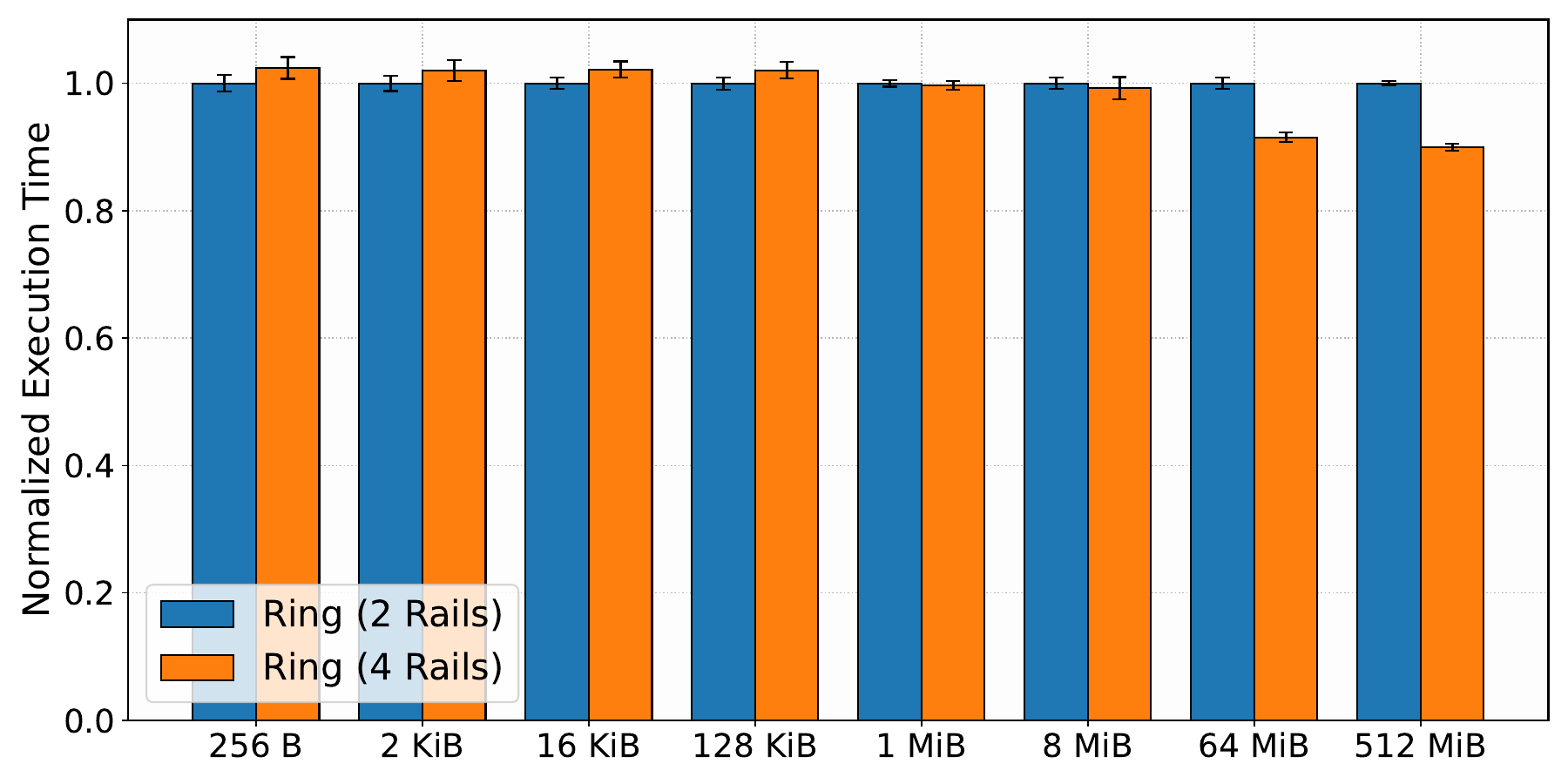}
  \caption{Ring \texttt{MPI\_Allreduce} on Leonardo (32 nodes; Open~MPI~4.1.6; UCX~1.15.0). Latency normalized to the default \texttt{UCX\_MAX\_RNDV\_RAILS=2} (lower is better). Setting \texttt{UCX\_MAX\_RNDV\_RAILS=4} (orange) yields up to a $10\%$ improvement over the default (blue).}
  \label{fig:rndv}
\end{figure}

Additionally, sub-optimality is not limited to algorithm choice: performance can change dramatically with backend/transport parameters that are easy to overlook if they are not made explicit in the experiment specification. To illustrate this sensitivity, we fix the \texttt{MPI\_Allreduce} algorithm to Ring on Leonardo at 32 nodes (removing algorithmic variability) and vary only \texttt{UCX\_MAX\_RNDV\_RAILS}, a UCX parameter that caps the number of network rails used by the rendezvous protocol for large-message transfers. Fig.~\ref{fig:rndv} reports execution times normalized to the default \texttt{UCX\_MAX\_RNDV\_RAILS=2}. For large messages in the rendezvous regime, increasing the rail limit to 4 reduces runtime up to $10\%$, whereas smaller messages (typically in the eager regime) are largely unaffected.

This result reinforces two practical points. First, meaningful tuning requires dealing with both algorithm selection and backend configuration (\textit{R3}): even a strong algorithm can appear weak under an unfavorable transport setting. Second, reproducibility and regression diagnosis depend on recording the \emph{effective} configuration used in each run (\textit{R5}). PICO captures both requested settings and platform configuration defaults, making it straightforward to rerun controlled A/B tests in which only a single knob changes and to interpret performance differences.

\begin{figure*}[!t]
\centering
\begin{minipage}[t]{0.66\textwidth}
  \centering
  \subfloat[Distance-Halving Binomial Tree]{\includegraphics[width=.49\linewidth]{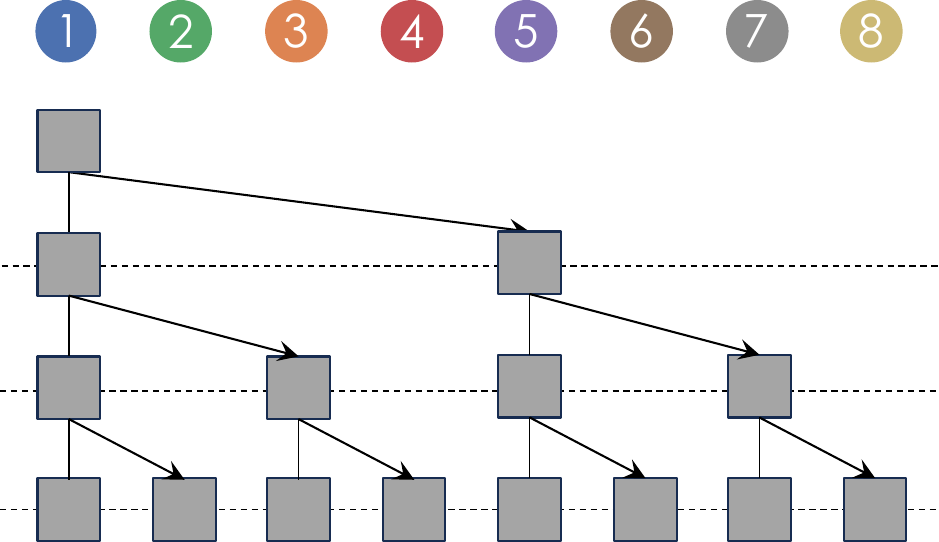}}%
  \hfill%
  \subfloat[Distance-Doubling Binomial Tree]{\includegraphics[width=.49\linewidth]{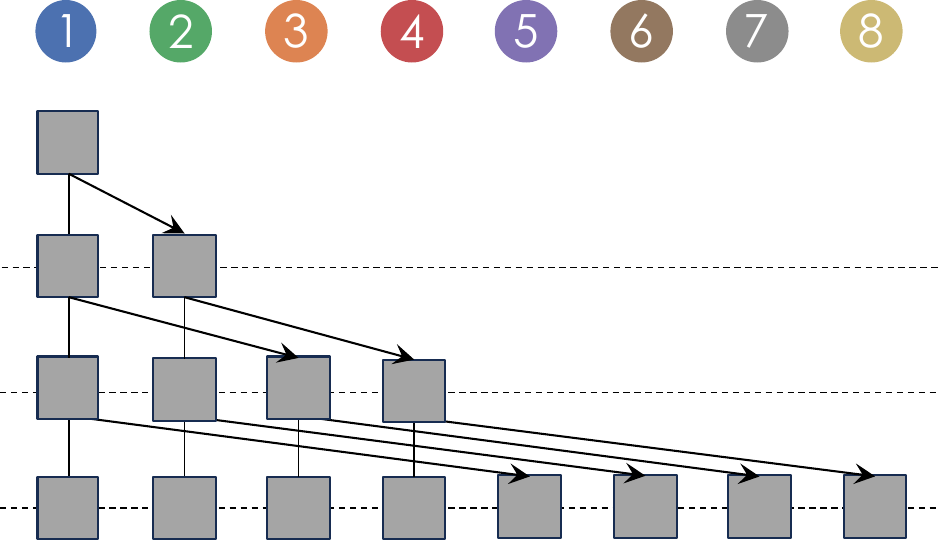}}
  \caption{Binomial-tree broadcast schedules with different partner ordering: (a) distance-halving vs.\ (b) distance-doubling. Both complete in $\log_2(p)$ rounds and transmit the same total volume, but differ in how communication distance evolves across steps. In particular, the distance halving approach maximizes communication locality at later communication rounds, when the overall communication volume is greater.} \label{fig:binomial-schedules}
\end{minipage}\hfill%
\begin{minipage}[t]{0.33\textwidth}
\centering
\includegraphics[width=0.83\linewidth]{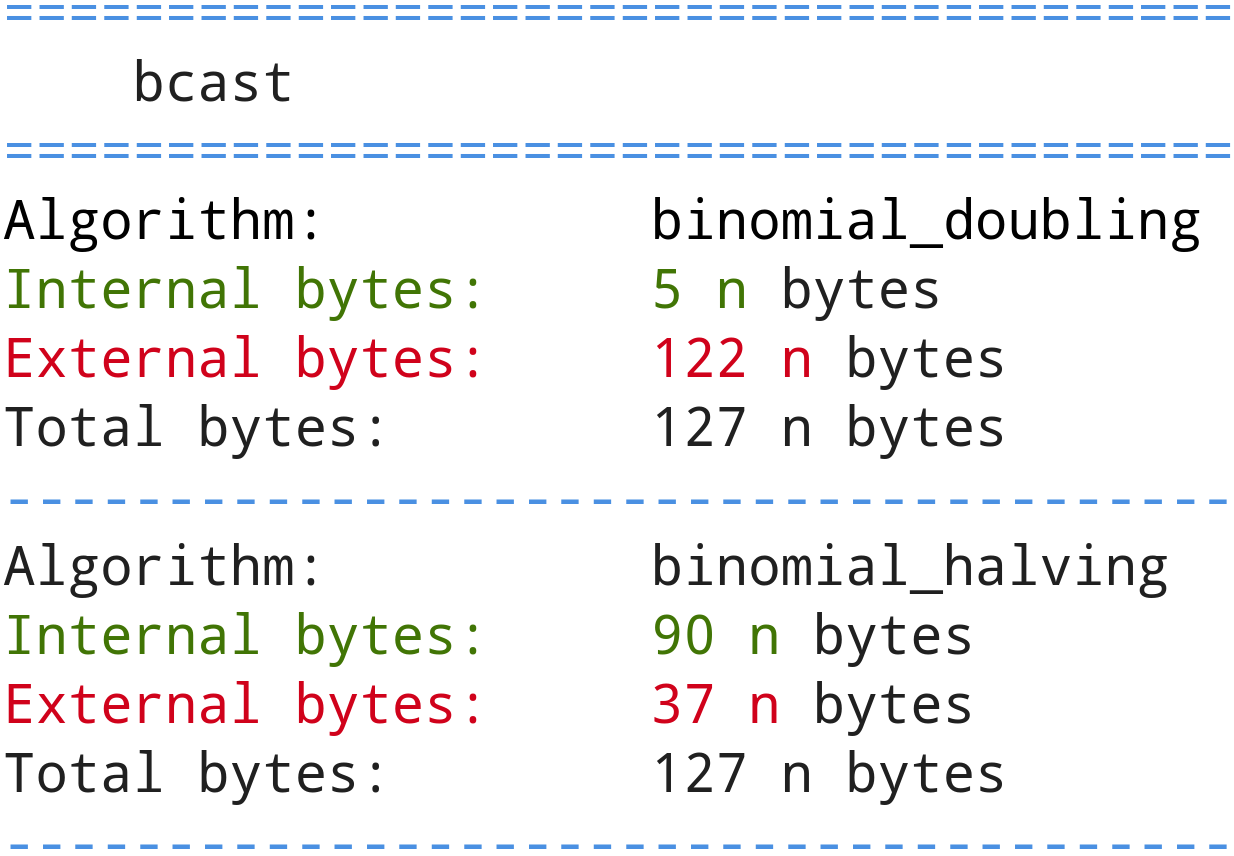}
\caption{Network volume estimates of distance-halving and distance-doubling broadcast on a 128 nodes allocation on Leonardo. Distance-halving broadcast induces only $29\%$ of total communication volume to across-group (\emph{external}) links, while distance-doubling $96\%$.}
\label{fig:bcast-tracer}
\end{minipage}
\end{figure*}

\subsection{Algorithmic differences}
\label{sec:case-algodiff}
Performance models are an essential tool for understanding collective behavior: they provide analytic guidance on how step count, overall communication volume, and reduction costs scale with process count and vector size. As discussed in Sec.~\ref{sec:requirements}, modern hierarchical systems can exhibit strong topology-dependent bottlenecks. Accurate performance prediction therefore benefits from refinements that track traffic flow and link saturation~\cite{uma2025peakalltoall, zhang2020congestion}. Our goal is not to argue against modeling, but to show why microbenchmarking remains a practical necessity: even when two algorithms are equivalent under a cost-model, their performance can differ substantially on different topologies and allocations.

We illustrate this by comparing two broadcast algorithms on Leonardo: \emph{distance-doubling} binomial tree broadcast (Open~MPI’s binomial broadcast implementation~\cite{openmpi_coll_base_bcast_dd6a7a3_L343}) and \emph{distance-halving} binomial tree broadcast (MPICH binomial tree implementation). The two communication schedules, illustrated in Fig.~\ref{fig:binomial-schedules}, appear indistinguishable under a classic $\alpha$--$\beta$ modeling: both complete in $\log_2(p)$ rounds and transmit the same total communication volume.
However, distance-doubling keeps communication local in the early rounds and defers longer-distance exchanges to later rounds. By contrast, distance-halving performs longer-distance exchanges earlier and becomes progressively more local in later rounds. On a topology with non-uniform link costs or tapered global bandwidth, this ordering can shift how many exchanges traverse local versus global links at each step, and thus where congestion pressure concentrates~\cite{desensi2024swing}.

\begin{figure}[!h]
\centering
\includegraphics[width=\linewidth]{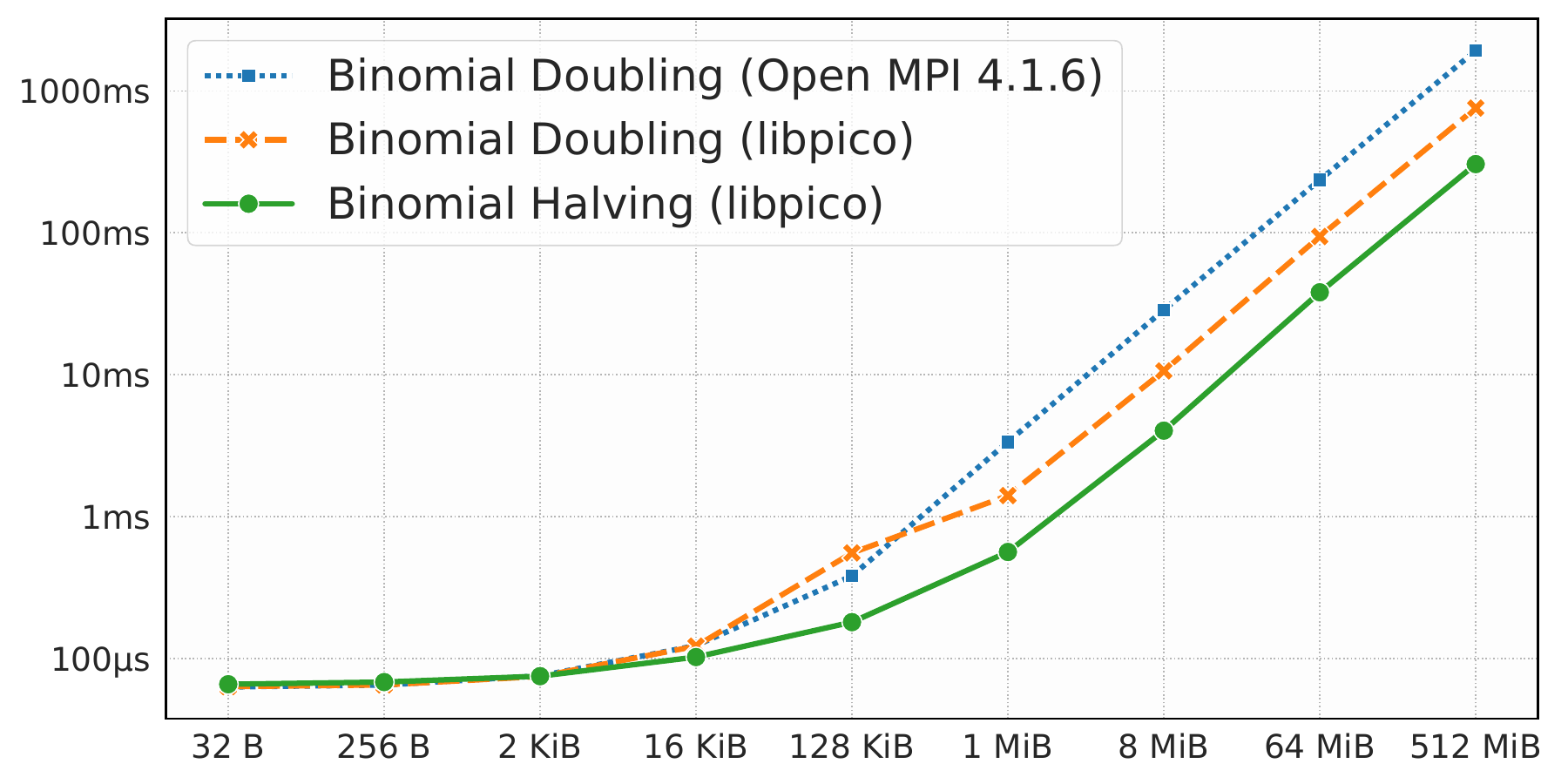}
\caption{Distance-doubling and distance-halving \texttt{MPI\_Bcast} on Leonardo, Open MPI (4.1.6). Latency vs.\ message size for 128 nodes, 4 processes per node (log--log axes); 512MiB: libpico 757ms (doubling) vs.\ 304ms (halving). Open MPI (doubling) 1,9s is one order of magnitude slower.}
\label{fig:bcast-results}
\end{figure}

To quantify this effect, PICO's network tracer reveals (Fig.~\ref{fig:bcast-tracer}) that, for the same 128 nodes allocation on Leonardo's Dragonfly network, distance-doubling algorithm sends nearly all volume inter-group (external 122$\cdot$n, internal 5$\cdot$n, where n is the size of the send buffer in Bytes), whereas distance-halving keeps 90$\cdot$n Bytes intra-group, reducing inter-group traffic to 37$\cdot$n. This effect arises naturally from the interaction of rank placement and algorithm’s communication schedule, illustrating why topology- and placement-aware diagnosis is valuable in realistic runs. Fig.~\ref{fig:bcast-results} reports the measured execution times of the two algorithms. The curves are nearly identical for small messages (up to 16\,KiB), but diverge sharply once large-message transfers dominate. At 512\,MiB, the distance-doubling algorithm is $\times2.5$ slower at 757\,ms compared to 304\,ms of the distance-halving one. Moreover, we can notice how the Open~MPI internal Binomial algorithm appears to be almost one order of magnitude slower at $1.9$\,s, indicating inefficiencies in its implementation regardless of the algorithm of choice.

Thus, cost models capture step/volume trade-offs, but they may not distinguish between algorithms that are equivalent in those metrics unless topology and placement effects are explicitly modeled. PICO complements modeling by enabling backend-neutral comparisons (\textit{R2}) while also providing placement-aware diagnosis thanks to its rich metadata gathering (\textit{R5}). In practice, these capabilities make it possible to identify when a default algorithm over-stresses links on hierarchical networks and to justify an alternative schedule with both structural evidence (tracer) and measured performance.

\begin{figure*}[!t]
  \centering
  \subfloat[Absolute time breakdown (log--log).\label{fig:instr-rab-line}]{
    \includegraphics[width=0.48\linewidth]{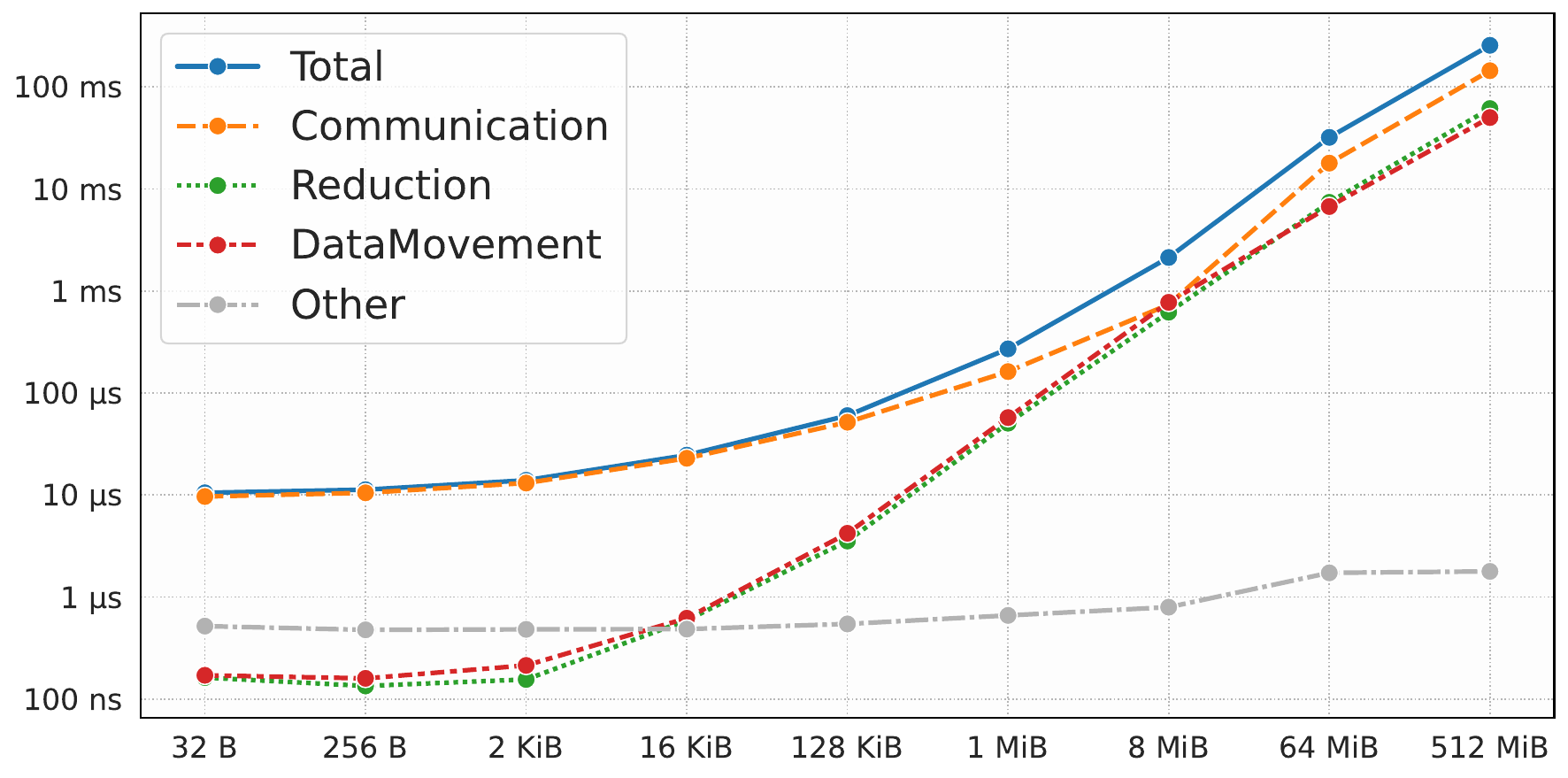}
  }%
  \hfill%
  \subfloat[Relative (percentage) breakdown.\label{fig:instr-rab-perc}]{
    \includegraphics[width=0.48\linewidth]{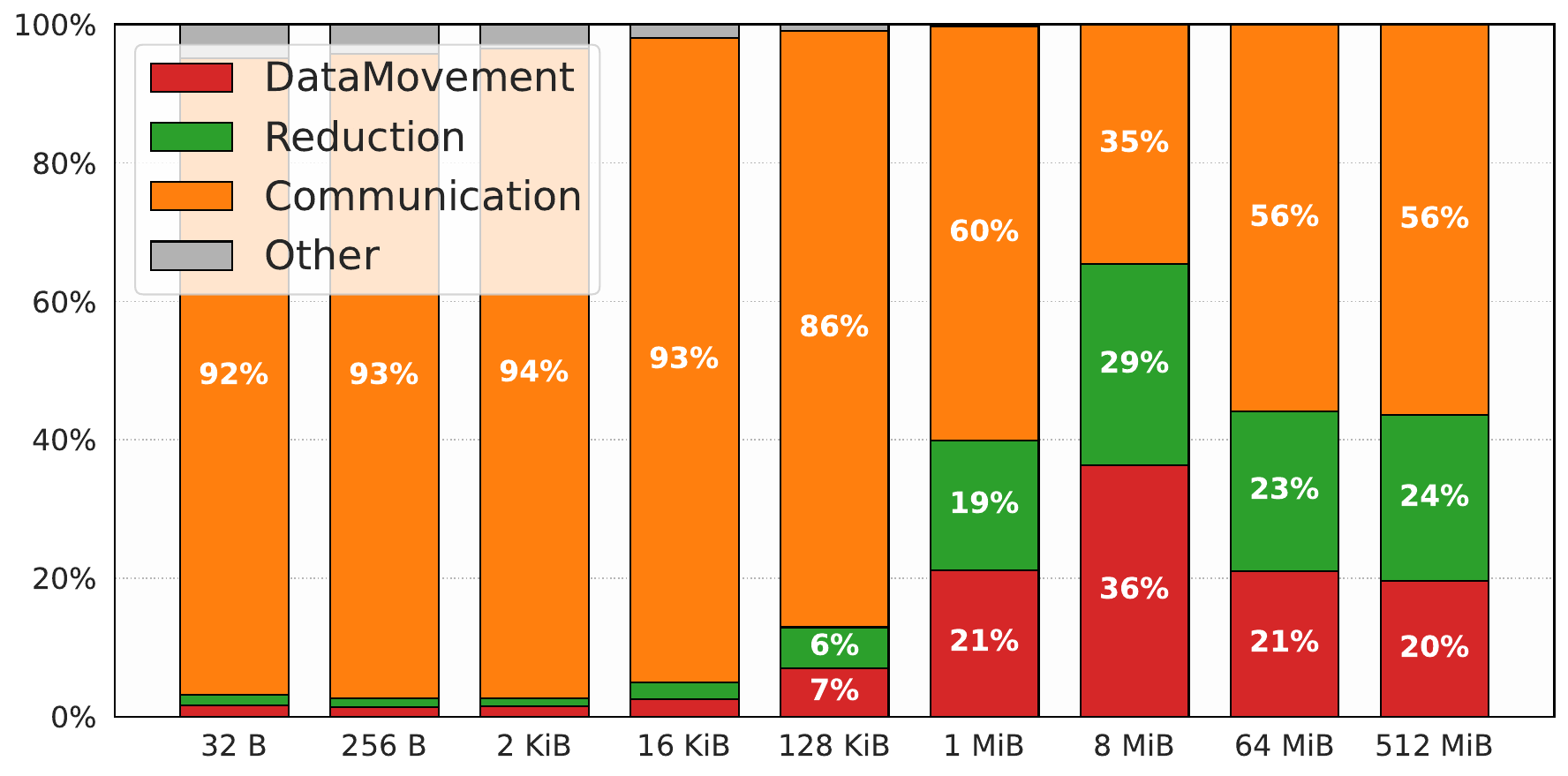}
  }
  \caption{Instrumented Rabenseifner Allreduce on 8 nodes (Leonardo, Open MPI 4.1.6; \texttt{libpico} implementation). (a) Absolute runtime breakdown into tagged components: Communication (network transfer), Reduction (compute), Data Movement (staging/copies to work buffers), and residual Other overhead. (b) Same data expressed as percentage shares, highlighting the shift from communication-dominated behavior at small messages to substantial data-movement and reduction contributions at larger messages.}
  \label{fig:instr-rab}
\end{figure*}

\subsection{Fine grained instrumentation} \label{sec:case-instr}
Collective communications are composed of multiple algorithmic steps, each one stressing different hardware resources (NIC/fabric, caches/DRAM and CPU/GPU arithmetic units). Crucially, the different performance characteristics of these resources shape how the cost of each step scales with message size and node count, meaning that the aggregate end-to-end timing of a collective can hide where inefficiencies originate.

An instrumented benchmark run of the \texttt{libpico} reference Rabenseifner implementation on an 8-node allocation on Leonardo (Fig.~\ref{fig:instr-rab}) reveals several non-obvious behaviors. Fig.~\ref{fig:instr-rab-line} shows the \emph{aggregate} runtime (blue) together with the tagged components—network communication (orange), reduction computation (green), and intra-node data movement (red; e.g., staging and copies to working buffers), while Fig.~\ref{fig:instr-rab-perc} reports their relative weight.

We first consider the overall trend in Fig.~\ref{fig:instr-rab-line}. For small message sizes (up to 128\,KiB), the communication curve is nearly indistinguishable from the aggregate curve, suggesting that end-to-end performance is dominated by network communication. In particular, for message sizes up to 2\,KiB the total runtime is nearly constant (10\,$\mu$s at 32\,B, 11\,$\mu$s at 256\,B, and 10\,$\mu$s at 2\,KiB), consistent with a latency-dominated regime where fixed network startup costs outweigh bandwidth effects. However, the tagged breakdown shows that this “communication-dominated” interpretation does not hold uniformly as message size grows. After 128\,KiB, Fig.~\ref{fig:instr-rab-perc} shows that the relative cost of communication drops sharply (from nearly $95\%$ to $35\%$) before increasing again to $56\%$ at 64\,MiB and 512\,MiB. This non-monotonic trend indicates that the scaling driver of the collective changes with message size: once per-message latency is amortized, the overall runtime is no longer governed by network transfer alone. The missing fraction is largely absorbed by \emph{intra-node data movement} and \emph{reduction}. As message size grows into the MiB range, the data-movement component rises substantially (staging and copies to work buffers), and reduction becomes a first-class contributor as more bytes must be combined at each step.

This breakdown can be interpreted through the intuition of the roofline model~\cite{williams2009roofline}: performance is limited by whichever resource is currently most constraining. While strictly speaking a collective communication is not a single kernel, its components can be interpreted as kernels with different limiting resources. The communication component is primarily limited by network latency for small messages and bandwidth for large ones, whereas the staging and reduction components are limited by local memory bandwidth and compute throughput (we assume local memory latency to be negligible). From this perspective, the non-monotonic behavior observed in Fig.~\ref{fig:instr-rab-line}--\ref{fig:instr-rab-perc} can be explained as follows: as message size grows, the dominant limiter shifts from network-latency to local data movement and reduction, effectively moving the collective onto a memory-bandwidth roof even though the operation is ``communication-heavy'' at a high level. At very large messages, the network bandwidth becomes the dominant limit, but the persistent contribution of data movement and reduction indicates that local memory bandwidth and compute throughput still cap end-to-end gains.

The crossover points at which this shift occurs, as well as the magnitude of the shift, depend on the machine’s hardware characteristics. Consequently, the same end-to-end Allreduce curve can hide different underlying bottlenecks on different systems, reinforcing the need for fine grained profiling when tuning or diagnosing performance. Thus, PICO’s instrumentation (\textit{R1}) on backend-neutral baselines (\textit{R2}) exposes hidden bottleneck shifts that are opaque under end-to-end timing.

\begin{figure*}[!h]
    \centering
    \includegraphics[width=\textwidth]{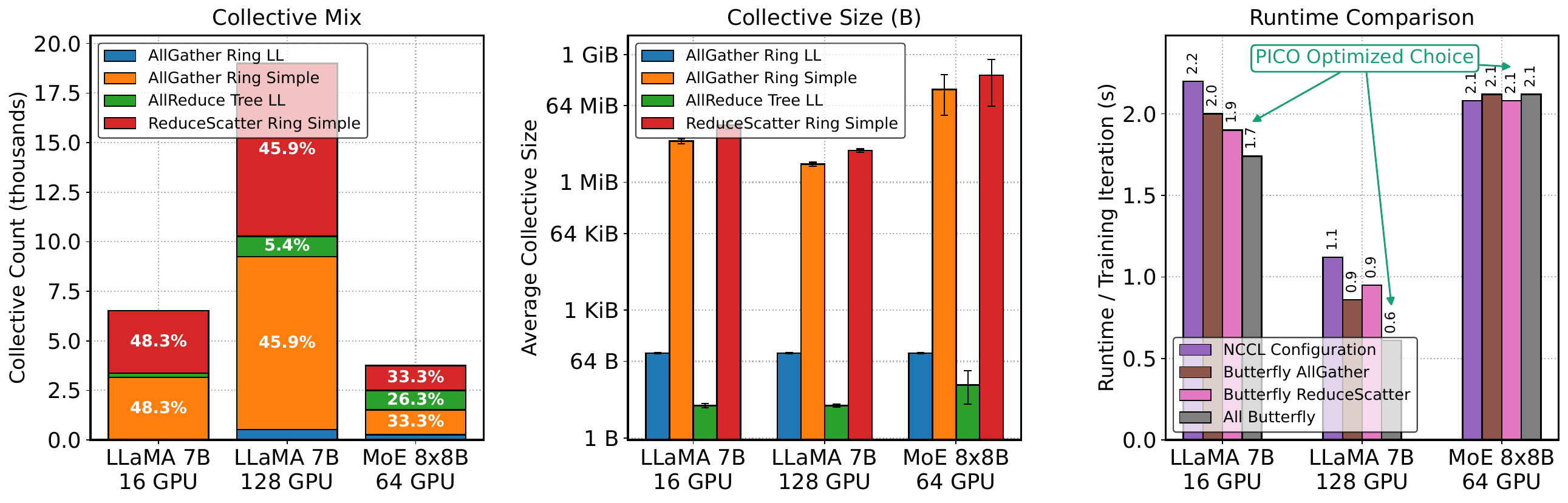}
    \caption{ATLAHS-based trace analysis and replay for AI training workloads. Left/center: collective mix and message-size distributions extracted from NCCL traces. Right: projected per-iteration time after substituting collective algorithms/protocols while preserving the original invocation sequence and message sizes. PICO-derived profiles reduce per-iteration time of up to $44\%$.}
    \label{fig:atlahs}
\end{figure*}

\subsection{Simulation results with ATLAHS} \label{sec:case-sim}
We evaluate PICO on real AI workloads using trace replay to translate microbenchmark-level effects into end-to-end performance changes. Specifically, we use the ATLAHS~\cite{bonato2025atlahs}, a recently published toolchain that can trace NCCL executions and replay them via GOAL traces~\cite{goal} running on a number of network simulators. ATLAHS can generate different replayable traces from the same raw NCCL logs, allowing us to swap collective algorithms and protocol choice while preserving original invocation sequence and message sizes. This enables controlled what-if analysis without needing to re-run the actual workload. Additionally, this capability allows us to test collective algorithms that might not be already implemented in NCCL by simply implementing new translation units in the simulator toolchain. The choice of application traces was intentionally geared toward AI workloads, as they feature large, repeated collectives for which algorithm, protocol, and transport decisions are more likely to produce observable end-to-end effects. Similar sensitivity can also arise in traditional HPC applications with communication-heavy phases, such as those based on 3D FFTs and their associated all-to-all exchanges~\cite{dalcin2018fastparallelmultidimensionalfft}; however, such traces were not available among the open-source traces considered in this study.

The traces we consider are collected with NCCL~2.22, a version that provides Ring and Tree implementations for AllReduce (i.e., Distance-Halving Reduce followed by Distance-Doubling Broadcast), but crucially only a Ring algorithm for ReduceScatter and AllGather. Newer NCCL releases have since added an additional algorithm, PAT~\cite{jeaugey2025patnewalgorithmallgather}, which is a Binomial Butterfly. During tracing, both the collective \emph{algorithm} and \emph{protocol} were recorded for each invocation, alongside several other key pieces of information such as the collective size and details about the communicator and GPU streams. While the algorithm controls the overall schedule of the collective, the protocol controls the low-level transfer/synchronization strategy; available NCCL protocols are \textit{Simple}, which favors large-message bandwidth, and \textit{LL}, which reduces small-message latency via flag-based synchronization~\cite{hu2025demystifyingncclindepthanalysis}.

Our chosen workloads are LLaMA 7B \cite{touvron2023llamaopenefficientfoundation}, run on 16 and 128 GPUs, and a Mistral MoE (Mixture-of-Experts) \cite{jiang2024mixtralexperts} on 64 GPUs: for clarity of exposition we will refer to those traces as \texttt{L16}, \texttt{L128} and \texttt{MoE}.
Analyzing number and types of collective invocations present in the raw traces we notice that (Fig.~\ref{fig:atlahs}; left) for both \texttt{L16}, and \texttt{L128} the overall majority of those collective are AllGather Ring Simple (\texttt{L16} 48.3\%, \texttt{L128} 45.9\%), and ReduceScatter Ring Simple (\texttt{L16} 48.3\%, \texttt{L128} 45.9\%), while the Allreduce Tree LL and ReduceScatter Ring LL accounted for a small minority of invocations (\texttt{L16} 1-3\%, \texttt{L128} 3-6\%). In the \texttt{MoE} trace we notice an overall lower number of invocations almost equally distributed between Allreduce Tree LL, ReduceScatter Ring Simple and Allgather Ring Simple. Taking into consideration size distribution (Fig.~\ref{fig:atlahs}; center) it appears that: (i) Allreduce invocations were all of relatively small size (i.e., $<1$\,KiB), (ii) AllGather and ReduceScatter invocations had a median size of 3-6\,MiB (\texttt{L16}) and 7-14\,MiB (\texttt{L128}), while (iii) a significantly higher size of 33-67\,MiB in the \texttt{MoE} trace. We note that for this analysis we purposely ignored point-to-point sends to focus purely on the collective operations.

Using PICO, we identified candidate \emph{collective profiles} (algorithm/protocol choices) for the observed communicator and message-size distributions. In particular, for the \texttt{L16} and \texttt{L128} traces we identified a profile consisting of AllGather and ReduceScatter Binomial Butterfly algorithm with Simple protocol, and Allreduce Tree algorithm with LL protocol.

Fig.~\ref{fig:atlahs} (right) reports the resulting simulated end-to-end runtimes, reported over a single training iteration: the PICO-optimized profiles improve over native NCCL by $21\%$ on \texttt{L16} and $44\%$ on \texttt{L128}. On the other hand, the optimized profile found for the MoE model reported no measurable improvements, indicating that a good profile was already in use when the trace was instrumented. This is likely because MoE has larger collectives on average, which tend to perform better with Ring implementations. For completeness, alternative suboptimal profiles were played alongside the optimal one, confirming runtime variation across algorithm/protocol choices, thus further highlighting the workload sensitivity to collective configurations.

Due to PICO's extensible design (\textit{R6}), we are able to evaluate NCCL algorithms and conduct a systematic evaluation campaign (\textit{R4}) across both algorithms selection and protocol configuration (\textit{R3}), observing end-to-end runtime improvements of real world applications.

\section{Conclusions}
We presented PICO, a lightweight and extensible framework for benchmarking collective communication operations across heterogeneous HPC and AI systems. Unlike existing tools, PICO integrates fine-grained profiling, rich metadata collection, and automated orchestration to support reproducible, system-aware performance analysis. Through its modular architecture, PICO enables portable comparisons across MPI, *CCL, and user-defined algorithms, while its integrated post-processing tools facilitate both high-level performance summaries and detailed algorithmic phase breakdowns.

Our case studies demonstrated how PICO can reveal suboptimal default algorithm selections and guide library tuning, highlight subtle performance trade-offs between closely related algorithms, quantify the impact of backend parameters, and identify hidden bottlenecks thanks to its reference backend-neutral implementations and instrumentation capabilities. Finally, we demonstrated that PICO tuned collective configurations can translate to real world end-to-end improvements.

These examples underscore PICO's utility for algorithm designers, application developers, and system administrators alike. By bridging systematic benchmarking with actionable insights, PICO aims to become a foundational tool for advancing the reproducible performance analysis of collective communications in next-generation computing systems.

\section{Acknowledgments}
This work is supported by the European Union’s Horizon Europe under grant 101175702 (NET4EXA), by Sapienza University Grants ADAGIO and D2QNeT (\textit{Bando per la ricerca di Ateneo} 2023 and 2024), and by a research grant from Microsoft Azure. The research was also conducted as part of the FastTrackAI project at the Singapore-ETH Centre, which was established collaboratively between ETH Zurich and the National Research Foundation, Singapore. Additionally, this research is supported by the National Research Foundation, Singapore (NRF), and the Ministry of Digital Development and Information (MDDI) under the AI Visiting Professorship (Award No. AIVP-2025-005). We acknowledge ISCRA for awarding this project access to the LEONARDO supercomputer, owned by the EuroHPC Joint Undertaking, hosted by CINECA (Italy). We acknowledge the EuroHPC Joint Undertaking, the LUMI consortium, and BSC for granting access to the LUMI and MareNostrum 5 supercomputers. These resources, hosted by CSC (Finland) and the Barcelona Supercomputing Center (Spain), were provided through the EuroHPC Regular Access program.

\printbibliography

@article{Liao2018,
  author    = {Xiang-ke Liao and Kai Lu and Can-qun Yang and Jin-wen Li and Yuan Yuan and Ming-che Lai and Li-bo Huang and Ping-jing Lu and Jian-bin Fang and Jing Ren and Jie Shen},
  title     = {Moving from exascale to zettascale computing: challenges and techniques},
  journal   = {Frontiers of Information Technology \& Electronic Engineering},
  volume    = {19},
  number    = {10},
  pages     = {1236--1244},
  year      = {2018},
  doi       = {10.1631/FITEE.1800494},
  url       = {https://doi.org/10.1631/FITEE.1800494},
  issn      = {2095-9230}
}

@misc{LLNL2025_ElCapitan,
  author       = {{Lawrence Livermore National Laboratory}},
  title        = {{Lawrence Livermore National Laboratory’s El Capitan verified as world’s fastest supercomputer}},
  howpublished = {Press release},
  month        = feb,
  year         = {2025},
  url          = {https://www.llnl.gov/article/52061/lawrence-livermore-national-laboratorys-el-capitan-verified-worlds-fastest-supercomputer},
}

@misc{top500_2025_06,
  author       = {{TOP500 Project}},
  title        = {{TOP500 List – June 2025}},
  year         = {2025},
  month        = jun,
  howpublished = {\url{https://top500.org/lists/top500/2025/06/}},
  note         = {Accessed: Jul. 18, 2025}
}

@TechReport{Exascale2014,
  author      = {Lucas, Robert and Ang, James and Bergman, Keren and Borkar, Shekhar and Carlson, William and Carrington, Laura and Chiu, George and Colwell, Robert and Dally, William and Dongarra, Jack and Geist, Al and Haring, Rud and Hittinger, Jeffrey and Hoisie, Adolfy and Klein, Dean Micron and Kogge, Peter and Lethin, Richard and Sarkar, Vivek and Schreiber, Robert and Shalf, John and Sterling, Thomas and Stevens, Rick and Bashor, Jon and Brightwell, Ron and Coteus, Paul and Debenedictus, Erik and Hiller, Jon and Kim, K. H. and Langston, Harper and Murphy, Richard Micron and Webster, Clayton and Wild, Stefan and Grider, Gary and Ross, Rob and Leyffer, Sven and Laros III, James},
  institution = {U.S.\ Department of Energy, ASCR},
  title       = {DOE Advanced Scientific Computing Advisory Subcommittee (ASCAC) Report: Top Ten Exascale Research Challenges},
  doi         = {10.2172/1222713},
  type        = {Report},
  year        = {2014},
}

@Article{Lu2022,
AUTHOR = {Lu, Ping-Jing and Lai, Ming-Che and Chang, Jun-Sheng},
TITLE = {A Survey of High-Performance Interconnection Networks in High-Performance Computer Systems},
JOURNAL = {Electronics},
VOLUME = {11},
YEAR = {2022},
NUMBER = {9},
ARTICLE-NUMBER = {1369},
URL = {https://www.mdpi.com/2079-9292/11/9/1369},
ISSN = {2079-9292},
DOI = {10.3390/electronics11091369}
}

@book{hwang1998scalable,
  title={Scalable Parallel Computing: Technology, Architecture, Programming},
  author={Hwang, K. and Xu, Z.},
  isbn={9780070317987},
  lccn={97041663},
  series={Computer engineering series},
  url={https://books.google.it/books?id=OJNQAAAAMAAJ},
  year={1998},
  publisher={WCB/McGraw-Hill}
}

@article{Weingram2023,
  author    = {Adam Weingram and Yuke Li and Hao Qi and Darren Ng and Liuyao Dai and Xiaoyi Lu},
  title     = {{xCCL: A Survey of Industry-Led Collective Communication Libraries for Deep Learning}},
  journal   = {Journal of Computer Science and Technology},
  volume    = {38},
  number    = {1},
  pages     = {166--195},
  year      = {2023},
  month     = feb,
  doi       = {10.1007/s11390-023-2894-6},
  url       = {https://doi.org/10.1007/s11390-023-2894-6},
  issn      = {1860-4749}
}

@misc{basu2024efficientalltoallcollectivecommunication,
      title={Efficient All-to-All Collective Communication Schedules for Direct-Connect Topologies}, 
      author={Prithwish Basu and Liangyu Zhao and Jason Fantl and Siddharth Pal and Arvind Krishnamurthy and Joud Khoury},
      year={2024},
      eprint={2309.13541},
      archivePrefix={arXiv},
      primaryClass={cs.DC},
      url={https://arxiv.org/abs/2309.13541}, 
}

@misc{desensi2024swing,
      title={Swing: Short-cutting Rings for Higher Bandwidth Allreduce}, 
      author={Daniele De Sensi and Tommaso Bonato and David Saam and Torsten Hoefler},
      year={2024},
      eprint={2401.09356},
      archivePrefix={arXiv},
      primaryClass={cs.DC},
      url={https://arxiv.org/abs/2401.09356}, 
}

@misc{bienz2022bruckallgather,
      title={A Locality-Aware Bruck Allgather}, 
      author={Amanda Bienz and Shreeman Gautam and Amun Kharel},
      year={2022},
      eprint={2206.03564},
      archivePrefix={arXiv},
      primaryClass={cs.DC},
      url={https://arxiv.org/abs/2206.03564}, 
}

@inproceedings{sewell2024bruck,
author = {Sewell, Andres and Fan, Ke and Shovon, Ahmedur Rahman and Dyken, Landon and Kumar, Sidharth and Petruzza, Steve},
title = {Bruck Algorithm Performance Analysis for Multi-GPU All-to-All Communication},
year = {2024},
isbn = {9798400708893},
publisher = {Association for Computing Machinery},
address = {New York, NY, USA},
url = {https://doi.org/10.1145/3635035.3635047},
doi = {10.1145/3635035.3635047},
booktitle = {Proceedings of the International Conference on High Performance Computing in Asia-Pacific Region},
pages = {127–133},
numpages = {7},
location = {Nagoya, Japan},
series = {HPCAsia '24}
}

@article{Sack2015collective,
author = {Sack, Paul and Gropp, William},
title = {Collective Algorithms for Multiported Torus Networks},
year = {2015},
issue_date = {January 2015},
publisher = {Association for Computing Machinery},
address = {New York, NY, USA},
volume = {1},
number = {2},
issn = {2329-4949},
url = {https://doi.org/10.1145/2686882},
doi = {10.1145/2686882},
journal = {ACM Trans. Parallel Comput.},
month = feb,
articleno = {12},
numpages = {33}
}

@INPROCEEDINGS{Dongkyun2025tidalmesh,
  author={Lim, Dongkyun and Kim, John},
  booktitle={2025 IEEE International Symposium on High Performance Computer Architecture (HPCA)}, 
  title={TidalMesh: Topology-Driven AllReduce Collective Communication for Mesh Topology}, 
  year={2025},
  volume={},
  number={},
  pages={1526-1540},
  doi={10.1109/HPCA61900.2025.00114}}

@inproceedings{laguna2019mpi_study,
author = {Laguna, Ignacio and Marshall, Ryan and Mohror, Kathryn and Ruefenacht, Martin and Skjellum, Anthony and Sultana, Nawrin},
title = {A large-scale study of MPI usage in open-source HPC applications},
year = {2019},
isbn = {9781450362290},
publisher = {Association for Computing Machinery},
address = {New York, NY, USA},
url = {https://doi.org/10.1145/3295500.3356176},
doi = {10.1145/3295500.3356176},
booktitle = {Proceedings of the International Conference for High Performance Computing, Networking, Storage and Analysis},
articleno = {31},
numpages = {14},
location = {Denver, Colorado},
series = {SC '19}
}

@inproceedings{balaji2009MPI_on_a_million,
author = {Balaji, Pavan and Buntinas, Darius and Goodell, David and Gropp, William and Kumar, Sameer and Lusk, Ewing and Thakur, Rajeev and Tr\"{a}ff, Jesper Larsson},
title = {MPI on a Million Processors},
year = {2009},
isbn = {9783642037696},
publisher = {Springer-Verlag},
address = {Berlin, Heidelberg},
url = {https://doi.org/10.1007/978-3-642-03770-2_9},
doi = {10.1007/978-3-642-03770-2_9},
booktitle = {Proceedings of the 16th European PVM/MPI Users' Group Meeting on Recent Advances in Parallel Virtual Machine and Message Passing Interface},
pages = {20–30},
numpages = {11},
location = {Espoo, Finland}
}

@misc{OSU,
  author       = {{The Ohio State University}},
  title        = {{OSU Micro-Benchmarks (OMB)}},
  howpublished = {Online},
  note         = {Accessed Jul. 21, 2025},
  url          = {https://mvapich.cse.ohio-state.edu/benchmarks/}
}

@misc{Intel_IMB_2021,
  author       = {{Intel Corporation}},
  title        = {{Intel® MPI Benchmarks User Guide, version 2021.2}},
  howpublished = {Online},
  year         = {2021},
  note         = {Accessed Jul. 21, 2025},
  url          = {https://www.intel.com/content/www/us/en/docs/mpi-library/user-guide-benchmarks/2021-2/overview.html}
}

@misc{nccl_test,
  author       = {{NVIDIA Corporation}},
  title        = {{NCCL-Tests}: Performance and correctness micro-benchmarks for NVIDIA NCCL},
  howpublished = {GitHub repository},
  note         = {Accessed Jul. 21, 2025},
  url          = {https://github.com/NVIDIA/nccl-tests}
}

@inproceedings{hidayetoglu2024commbench,
  title={CommBench: Micro-Benchmarking Hierarchical Networks with Multi-GPU, Multi-NIC Nodes},
  author={Hidayetoglu, Mert and De Gonzalo, Simon Garcia and Slaughter, Elliott and Li, Yu and Zimmer, Christopher and Bicer, Tekin and Ren, Bin and Gropp, William and Hwu, Wen-Mei and Aiken, Alex},
  booktitle={Proceedings of the 38th ACM International Conference on Supercomputing},
  pages={426--436},
  year={2024}
}

@misc{nccl,
  author       = {{NVIDIA Corporation}},
  title        = {{NVIDIA Collective Communication Library (NCCL) Documentation}},
  howpublished = {Online documentation},
  year         = {2025},
  note         = {Accessed July 23, 2025},
  url          = {https://docs.nvidia.com/deeplearning/nccl/index.html}
}

@INPROCEEDINGS{accl,
  author={He, Zhenhao and Parravicini, Daniele and Petrica, Lucian and O’Brien, Kenneth and Alonso, Gustavo and Blott, Michaela},
  booktitle={2021 IEEE/ACM International Workshop on Heterogeneous High-performance Reconfigurable Computing (H2RC)},
  title={ACCL: FPGA-Accelerated Collectives over 100 Gbps TCP-IP},
  year={2021},
  pages={33-43},
  doi={10.1109/H2RC54759.2021.00009}}

@misc{rccl,
  author       = {{Advanced Micro Devices, Inc.}},
  title        = {{ROCm Communication Collectives Library (RCCL) Documentation, v2.22.3}},
  howpublished = {Online documentation},
  year         = {2025},
  note         = {Accessed Jul. 23, 2025},
  url          = {https://rocm.docs.amd.com/projects/rccl}
}

@misc{openucx-website,
    title = {{The Unified Communication X Library}},
    key = {{The Unified Communication X Library}},
    howpublished = {{\url{http://www.openucx.org}}}
}

@misc{libfabric,
  author       = {{OpenFabrics Interfaces Working Group (OFIWG)}},
  title        = {{libfabric}: Open Fabric Interfaces framework for high-performance networking},
  howpublished = {GitHub repository},
  year         = {2025},
  note         = {Accessed Jul. 23, 2025},
  url          = {https://github.com/ofiwg/libfabric}
}

@article{qureshi2020resource_allocation,
author = {Muhammad Shuaib Qureshi and Muhammad Bilal Qureshi and Muhammad Fayaz and Wali Khan Mashwani and Samir Brahim Belhaouari and Saima Hassan and Asadullah Shah},
title ={A comparative analysis of resource allocation schemes for real-time services in high-performance computing systems},
journal = {International Journal of Distributed Sensor Networks},
volume = {16},
number = {8},
pages = {1550147720932750},
year = {2020},
doi = {10.1177/1550147720932750},
URL = { https://doi.org/10.1177/1550147720932750 },
eprint = {    https://doi.org/10.1177/1550147720932750 }
}

@InProceedings{andrii2017taskmapping,
author="Kovalov, Andrii and Lobe, Elisabeth and Gerndt, Andreas and L{\"u}dtke, Daniel",
editor="Polikarpova, Nadia and Schneider, Steve",
title="Task-Node Mapping in an Arbitrary Computer Network Using SMT Solver",
booktitle="Integrated Formal Methods",
year="2017",
publisher="Springer International Publishing",
address="Cham",
pages="177--191",
isbn="978-3-319-66845-1"
}

@INPROCEEDINGS{zhang2020congestion,
  author={Zhang, Yijia and Groves, Taylor and Cook, Brandon and Wright, Nicholas J. and Coskun, Ayse K.},
  booktitle={2020 IEEE International Conference on Cluster Computing (CLUSTER)}, 
  title={Quantifying the impact of network congestion on application performance and network metrics}, 
  year={2020},
  volume={},
  number={},
  pages={162-168},
  doi={10.1109/CLUSTER49012.2020.00026}}

@inproceedings{gpugpuinterconnect,
  author = {De Sensi, Daniele and Pichetti, Lorenzo and Vella, Flavio and De Matteis, Tiziano and Ren, Zebin and Fusco, Luigi and Turisini, Matteo and Cesarini, Daniele and Lust, Kurt and Trivedi, Animesh and Roweth, Duncan and Spiga, Filippo and Di Girolamo, Salvatore and Hoefler, Torsten},
  title = {Exploring GPU-to-GPU Communication: Insights into Supercomputer Interconnects},
  year = {2024},
  month = nov,
  booktitle = {Proceedings of the International Conference for High Performance Computing, Networking, Storage and Analysis (SC'24)},
  doi = {10.1109/SC41406.2024.00039},
  dimensions = {true},
}

@misc{hu2025demystifyingncclindepthanalysis,
      title={Demystifying NCCL: An In-depth Analysis of GPU Communication Protocols and Algorithms}, 
      author={Zhiyi Hu and Siyuan Shen and Tommaso Bonato and Sylvain Jeaugey and Cedell Alexander and Eric Spada and James Dinan and Jeff Hammond and Torsten Hoefler},
      year={2025},
      eprint={2507.04786},
      archivePrefix={arXiv},
      primaryClass={cs.DC},
      url={https://arxiv.org/abs/2507.04786}, 
}

@misc{NVIDIA_NVLink_Blog_2023,
  author       = {Rick Merritt},
  title        = {What Is NVLink?},
  howpublished = {NVIDIA Official Blog},
  month        = mar,
  year         = {2023},
  note         = {Accessed August 8, 2025},
  url          = {https://blogs.nvidia.com/blog/what-is-nvidia-nvlink/}
}

@misc{Schor2018Zeppelin,
  author       = {David Schor},
  title        = {{ISSCC 2018: AMD’s Zeppelin; Multi-chip routing and packaging}},
  howpublished = {WikiChip Fuse blog},
  month        = mar,
  year         = {2018},
  note         = {Accessed Aug. 11, 2025},
  url          = {https://fuse.wikichip.org/news/1064/isscc-2018-amds-zeppelin-multi-chip-routing-and-packaging/}
}

@misc{Synopsys2025UltraEthernetUALink,
  author       = {Jon Ames and Ron Lowman},
  title        = {{How Ultra Ethernet and UALink Enable High-Performance, Scalable AI Networks}},
  howpublished = {Synopsys Blog},
  month        = jan,
  year         = {2025},
  note         = {Accessed Aug. 11, 2025},
  url          = {https://www.synopsys.com/articles/ultra-ethernet-ualink-ai-networks.html}
}

@misc{Broadcom_SUE_Spec_2025,
  author       = {{Broadcom Inc.}},
  title        = {{Scale-Up Ethernet (SUE) Framework Specification}},
  howpublished = {Technical specification (PDF)},
  month        = jul,
  year         = {2025},
  note         = {Accessed Aug. 11, 2025},
  url          = {https://docs.broadcom.com/doc/scale-up-ethernet-framework}
}

@INBOOK{Buyya_2002_infiniband,
  author={Buyya, Rajkumar and Cortes, Toni and Jin, Hai},
  booktitle={High Performance Mass Storage and Parallel I/O: Technologies and Applications}, 
  title={An Introduction to the InfiniBand Architecture}, 
  year={2002},
  volume={},
  number={},
  pages={616-632},
  doi={10.1109/9780470544839.ch42}}

@inproceedings{De_Sensi_2020_slingshot,
   title={An In-Depth Analysis of the Slingshot Interconnect},
   url={http://dx.doi.org/10.1109/SC41405.2020.00039},
   DOI={10.1109/sc41405.2020.00039},
   booktitle={SC20: International Conference for High Performance Computing, Networking, Storage and Analysis},
   publisher={IEEE},
   author={De Sensi, Daniele and Di Girolamo, Salvatore and McMahon, Kim H. and Roweth, Duncan and Hoefler, Torsten},
   year={2020},
   month=nov, pages={1–14} }

@misc{hoefler2025ultraethernetsdesignprinciples,
      title={Ultra Ethernet's Design Principles and Architectural Innovations}, 
      author={Torsten Hoefler and Karen Schramm and Eric Spada and Keith Underwood and Cedell Alexander and Bob Alverson and Paul Bottorff and Adrian Caulfield and Mark Handley and Cathy Huang and Costin Raiciu and Abdul Kabbani and Eugene Opsasnick and Rong Pan and Adee Ran and Rip Sohan},
      year={2025},
      eprint={2508.08906},
      archivePrefix={arXiv},
      primaryClass={cs.NI},
      url={https://arxiv.org/abs/2508.08906}, 
}

@misc{NVIDIA_GB200_NVL72_2025,
  author       = {{NVIDIA Corporation}},
  title        = {{NVIDIA GB200 NVL72 (Grace + Blackwell) – Rack-Scale AI System}},
  howpublished = {Product web page},
  year         = {2025},
  note         = {Accessed Aug. 11, 2025},
  url          = {https://www.nvidia.com/en-us/data-center/gb200-nvl72/}
}

@misc{zuo2025servinglargelanguagemodels,
      title={Serving Large Language Models on Huawei CloudMatrix384}, 
      author={Pengfei Zuo and Huimin Lin and Junbo Deng and Nan Zou and Xingkun Yang and Yingyu Diao and Weifeng Gao and Ke Xu and Zhangyu Chen and Shirui Lu and Zhao Qiu and Peiyang Li and Xianyu Chang and Zhengzhong Yu and Fangzheng Miao and Jia Zheng and Ying Li and Yuan Feng and Bei Wang and Zaijian Zong and Mosong Zhou and Wenli Zhou and Houjiang Chen and Xingyu Liao and Yipeng Li and Wenxiao Zhang and Ping Zhu and Yinggang Wang and Chuanjie Xiao and Depeng Liang and Dong Cao and Juncheng Liu and Yongqiang Yang and Xiaolong Bai and Yi Li and Huaguo Xie and Huatao Wu and Zhibin Yu and Lv Chen and Hu Liu and Yujun Ding and Haipei Zhu and Jing Xia and Yi Xiong and Zhou Yu and Heng Liao},
      year={2025},
      eprint={2506.12708},
      archivePrefix={arXiv},
      primaryClass={cs.DC},
      url={https://arxiv.org/abs/2506.12708}, 
}

@inproceedings{ucc,
  author       = {Manjunath Gorentla Venkata and
                  Valentine Petrov and
                  Sergey Lebedev and
                  Devendar Bureddy and
                  Ferrol Aderholdt and
                  Joshua Ladd and
                  Gil Bloch and
                  Mike Dubman and
                  Gilad Shainer},
  title        = {Unified Collective Communication {(UCC):} An Unified Library for CPU,
                  GPU, and {DPU} Collectives},
  booktitle    = {{IEEE} Symposium on High-Performance Interconnects, {HOTI} 2024, Albuquerque,
                  NM, USA, August 21-23, 2024},
  pages        = {37--46},
  publisher    = {{IEEE}},
  year         = {2024},
  url          = {https://doi.org/10.1109/HOTI63208.2024.00018},
  doi          = {10.1109/HOTI63208.2024.00018},
  bibsource    = {dblp computer science bibliography, https://dblp.org}
}

@inproceedings{bine,
  author = {De Sensi, Daniele and Pasqualoni, Saverio and Piarulli, Lorenzo and Bonato, Tommaso and Ba, Seydou and Turisini, Matteo and Domke, Jens and Hoefler, Torsten},
  title = {Bine Trees: Enhancing Collective Operations by Optimizing Communication Locality},
  year = {2025},
  month = nov,
  booktitle = {Proceedings of the International Conference for High Performance Computing, Networking, Storage and Analysis (SC'25)},
  doi = {To Appear},
  dimensions = {true},
}

@inproceedings{frontier,
author = {Atchley, Scott and Zimmer, Christopher and Lange, John and Bernholdt, David and Melesse Vergara, Veronica and Beck, Thomas and Brim, Michael and Budiardja, Reuben and Chandrasekaran, Sunita and Eisenbach, Markus and Evans, Thomas and Ezell, Matthew and Frontiere, Nicholas and Georgiadou, Antigoni and Glenski, Joe and Grete, Philipp and Hamilton, Steven and Holmen, John and Huebl, Axel and Jacobson, Daniel and Joubert, Wayne and Mcmahon, Kim and Merzari, Elia and Moore, Stan and Myers, Andrew and Nichols, Stephen and Oral, Sarp and Papatheodore, Thomas and Perez, Danny and Rogers, David M. and Schneider, Evan and Vay, Jean-Luc and Yeung, P. K.},
title = {Frontier: Exploring Exascale},
year = {2023},
isbn = {9798400701092},
publisher = {Association for Computing Machinery},
address = {New York, NY, USA},
url = {https://doi.org/10.1145/3581784.3607089},
doi = {10.1145/3581784.3607089},
booktitle = {Proceedings of the International Conference for High Performance Computing, Networking, Storage and Analysis},
articleno = {52},
numpages = {16},
location = {Denver, CO, USA},
series = {SC '23}
}

@inproceedings{Zwinger2023LUMI,
  author    = {T. Zwinger and J. Heikonen and P. Manninen},
  title     = {LUMI supercomputer for European researchers},
  booktitle = {Galileo Conference: Solid Earth and Geohazards in the Exascale Era},
  address   = {Barcelona, Spain},
  date      = {2023-05-23/2023-05-26},
  pages     = {GC11-solidearth-25},
  year      = {2023},
  doi       = {10.5194/egusphere-gc11-solidearth-25},
  url       = {https://doi.org/10.5194/egusphere-gc11-solidearth-25}
}

@misc{turisini2023leonardopaneuropeanpreexascalesupercomputer,
      title={LEONARDO: A Pan-European Pre-Exascale Supercomputer for HPC and AI Applications}, 
      author={Matteo Turisini and Giorgio Amati and Mirko Cestari},
      year={2023},
      eprint={2307.16885},
      archivePrefix={arXiv},
      primaryClass={cs.DC},
      url={https://arxiv.org/abs/2307.16885}, 
}

@inproceedings{kim_2008_dragonfly,
author = {Kim, Jongryoul and Dally, William and Scott, Steve and Abts, Dennis},
year = {2008},
month = {07},
pages = {77-88},
title = {Technology-Driven, Highly-Scalable Dragonfly Topology},
volume = {36},
isbn = {978-0-7695-3174-8},
journal = {ACM SIGARCH Computer Architecture News},
doi = {10.1109/ISCA.2008.19}
}

@inproceedings{shpiner_2017_dragonflyplus,
author = {Shpiner, Alexander and Haramaty, Zachy and Eliad, Saar and Zdornov, Vladimir and Gafni, Barak and Zahavi, Eitan},
year = {2017},
month = {02},
pages = {},
title = {Dragonfly+: Low Cost Topology for Scaling Datacenters},
doi = {10.1109/HiPINEB.2017.11}
}

@misc{banchelli2025introducingmarenostrum5europeanpreexascale,
      title={Introducing MareNostrum5: A European pre-exascale energy-efficient system designed to serve a broad spectrum of scientific workloads}, 
      author={Fabio Banchelli and Marta Garcia-Gasulla and Filippo Mantovani and Joan Vinyals and Josep Pocurull and David Vicente and Beatriz Eguzkitza and Flavio C. C. Galeazzo and Mario C. Acosta and Sergi Girona},
      year={2025},
      eprint={2503.09917},
      archivePrefix={arXiv},
      primaryClass={cs.DC},
      url={https://arxiv.org/abs/2503.09917}, 
}

@misc{desensi2022noisecloudsinfluencenetwork,
      title={Noise in the Clouds: Influence of Network Performance Variability on Application Scalability}, 
      author={Daniele De Sensi and Tiziano De Matteis and Konstantin Taranov and Salvatore Di Girolamo and Tobias Rahn and Torsten Hoefler},
      year={2022},
      eprint={2210.15315},
      archivePrefix={arXiv},
      primaryClass={cs.DC},
      url={https://arxiv.org/abs/2210.15315}, 
}

@misc{Juelich2025JUPITER_Tech,
  author       = {{Forschungszentrum Jülich, Jülich Supercomputing Centre}},
  title        = {{JUPITER Technical Overview}},
  howpublished = {Technical overview page},
  year         = {2025},
  note         = {Last modified Jan 7, 2025; accessed Aug 11, 2025},
  url          = {https://www.fz-juelich.de/en/ias/jsc/jupiter/tech}
}

@INPROCEEDINGS{kandalla_2009_multi_ladder_allgather,
  author={Kandalla, Krishna and Subramoni, Hari and Santhanaraman, Gopal and Koop, Matthew and Panda, Dhabaleswar K.},
  booktitle={2009 IEEE International Symposium on Parallel \& Distributed Processing}, 
  title={Designing multi-leader-based Allgather algorithms for multi-core clusters}, 
  year={2009},
  volume={},
  number={},
  pages={1-8},
  doi={10.1109/IPDPS.2009.5160896}}

@InProceedings{larsson_2006_smp,
author="Tr{\"a}ff, Jesper Larsson",
editor="Mohr, Bernd
and Tr{\"a}ff, Jesper Larsson
and Worringen, Joachim
and Dongarra, Jack",
title="Efficient Allgather for Regular SMP-Clusters",
booktitle="Recent Advances in Parallel Virtual Machine and Message Passing Interface",
year="2006",
publisher="Springer Berlin Heidelberg",
address="Berlin, Heidelberg",
pages="58--65",
isbn="978-3-540-39112-8"
}

@misc{OpenMPI_CollTuned_5.0,
  author       = {{The Open MPI Community}},
  title        = {{Open MPI 5.0: 11.10. Tuning Collectives (coll-tuned)}},
  howpublished = {Online documentation},
  year         = {2025},
  note         = {Last updated Jul. 31, 2025; accessed Aug. 11, 2025},
  url          = {https://docs.open-mpi.org/en/v5.0.x/tuning-apps/coll-tuned.html}
}

@misc{jeaugey2025patnewalgorithmallgather,
      title={PAT: a new algorithm for all-gather and reduce-scatter operations at scale}, 
      author={Sylvain Jeaugey},
      year={2025},
      eprint={2506.20252},
      archivePrefix={arXiv},
      primaryClass={cs.DC},
      url={https://arxiv.org/abs/2506.20252}, 
}

@inproceedings{bonato2025atlahs,
author = {Shen, Siyuan and Bonato, Tommaso and Hu, Zhiyi and Jordan, Pasquale and Chen, Tiancheng and Hoefler, Torsten},
title = {ATLAHS: An Application-centric Network Simulator Toolchain for AI, HPC, and Distributed Storage},
year = {2025},
isbn = {9798400714665},
publisher = {Association for Computing Machinery},
address = {New York, NY, USA},
url = {https://doi.org/10.1145/3712285.3759838},
doi = {10.1145/3712285.3759838},
booktitle = {Proceedings of the International Conference for High Performance Computing, Networking, Storage and Analysis},
pages = {349–367},
numpages = {19},
location = {
},
series = {SC '25}
}

@INPROCEEDINGS{goal,
  author={Hoefler, Torsten and Siebert, Christian and Lumsdaine, Andrew},
  booktitle={2009 International Conference on Parallel Processing}, 
  title={Group Operation Assembly Language - A Flexible Way to Express Collective Communication}, 
  year={2009},
  volume={},
  number={},
  pages={574-581},
  doi={10.1109/ICPP.2009.70}}

@misc{nvidia2025ncclinspector,
  title        = {Enhancing Communication Observability of AI Workloads with NCCL Inspector},
  author       = {Sirshak Das and Jason Sewall and Giuseppe Congiu and Pavel Shamis and Gargi Prasad},
  year         = {2025},
  month        = dec,
  day          = {10},
  howpublished = {\url{https://developer.nvidia.com/blog/enhancing-communication-observability-of-ai-workloads-with-nccl-inspector/}},
  note         = {NVIDIA Developer Blog},
}

@inproceedings{uma2025peakalltoall,
author = {Uma-Vaideswaran, Rohini and Romero, Joshua and Dotson, Daniel L. and Appelhans, David and Yeung, P. K.},
title = {A Peak Performance Model for All-to-all on Hierarchical Systems and Its Applications},
year = {2025},
isbn = {9798400718717},
publisher = {Association for Computing Machinery},
address = {New York, NY, USA},
url = {https://doi.org/10.1145/3731599.3767704},
doi = {10.1145/3731599.3767704},
booktitle = {Proceedings of the SC '25 Workshops of the International Conference for High Performance Computing, Networking, Storage and Analysis},
pages = {1442–1451},
numpages = {10},
location = {
},
series = {SC Workshops '25}
}

@misc{openmpi_coll_base_bcast_dd6a7a3_L343,
  author       = {{Open MPI Project}},
  title        = {{ompi/mca/coll/base/coll\_base\_bcast.c}},
  howpublished = {\url{https://github.com/open-mpi/ompi/blob/dd6a7a3ad0c37dde58da7ccabe2c54da8f51d130/ompi/mca/coll/base/coll_base_bcast.c}},
  note         = {Commit dd6a7a3ad0c37dde58da7ccabe2c54da8f51d130, line 343. Accessed 2025-12-16},
  year         = {2025}
}

@inproceedings{wu2024mscc,
author = {Wu, Yongji and Xu, Yechen and Chen, Jingrong and Wang, Zhaodong and Zhang, Ying and Lentz, Matthew and Zhuo, Danyang},
title = {MCCS: A Service-based Approach to Collective Communication for Multi-Tenant Cloud},
year = {2024},
isbn = {9798400706141},
publisher = {Association for Computing Machinery},
address = {New York, NY, USA},
url = {https://doi.org/10.1145/3651890.3672252},
doi = {10.1145/3651890.3672252},
booktitle = {Proceedings of the ACM SIGCOMM 2024 Conference},
pages = {679–690},
numpages = {12},
location = {Sydney, NSW, Australia},
series = {ACM SIGCOMM '24}
}

@article{thakur2005optimization,
  title={Optimization of collective communication operations in MPICH},
  author={Thakur, Rajeev and Rabenseifner, Rolf and Gropp, William},
  journal={The International Journal of High Performance Computing Applications},
  volume={19},
  number={1},
  pages={49--66},
  year={2005},
  publisher={Sage Publications Sage CA: Thousand Oaks, CA}
}

@inproceedings{hoefler2007netgauge,
  title={Netgauge: A network performance measurement framework},
  author={Hoefler, Torsten and Mehlan, Torsten and Lumsdaine, Andrew and Rehm, Wolfgang},
  booktitle={International Conference on High Performance Computing and Communications},
  pages={659--671},
  year={2007},
  organization={Springer}
}

@inproceedings{hunold2014reproducibleMPImicro,
author = {Hunold, Sascha and Carpen-Amarie, Alexandra and Tr\"{a}ff, Jesper Larsson},
title = {Reproducible MPI Micro-Benchmarking Isn't As Easy As You Think},
year = {2014},
isbn = {9781450328753},
publisher = {Association for Computing Machinery},
address = {New York, NY, USA},
url = {https://doi.org/10.1145/2642769.2642785},
doi = {10.1145/2642769.2642785},
booktitle = {Proceedings of the 21st European MPI Users' Group Meeting},
pages = {69–76},
numpages = {8},
location = {Kyoto, Japan},
series = {EuroMPI/ASIA '14}
}

@article{hoefler-collmea,
  author={Torsten Hoefler and Timo Schneider and Andrew Lumsdaine},
  title={{Accurately Measuring Overhead, Communication Time and Progression of Blocking and Nonblocking Collective Operations at Massive Scale}},
  journal={International Journal of Parallel, Emergent and Distributed Systems},
  year={2010},
  month={Jul.},
  pages={241-258},
  volume={25},
  number={4},
  publisher={Taylor \& Francis Group},
  issn={1744-5779},
  source={http://www.unixer.de/~htor/publications/},
}

@inproceedings{hunold2015impact,
  title={On the impact of synchronizing clocks and processes on benchmarking MPI collectives},
  author={Hunold, Sascha and Carpen-Amarie, Alexandra},
  booktitle={Proceedings of the 22nd European MPI Users' Group Meeting},
  pages={1--10},
  year={2015}
}

@misc{hunsa_reprompi_2025,
  author       = {Hunold Sascha},
  title        = {ReproMPI Benchmark for MPI Collective},
  year         = {2025},
  howpublished = {\url{https://github.com/hunsa/reprompi}},
  note         = {GitHub repository. Accessed: 2025-12-19.}
}

@article{williams2009roofline,
  title={Roofline: an insightful visual performance model for multicore architectures},
  author={Williams, Samuel and Waterman, Andrew and Patterson, David},
  journal={Communications of the ACM},
  volume={52},
  number={4},
  pages={65--76},
  year={2009},
  publisher={ACM New York, NY, USA}
}

@misc{touvron2023llamaopenefficientfoundation,
      title={LLaMA: Open and Efficient Foundation Language Models}, 
      author={Hugo Touvron and Thibaut Lavril and Gautier Izacard and Xavier Martinet and Marie-Anne Lachaux and Timothée Lacroix and Baptiste Rozière and Naman Goyal and Eric Hambro and Faisal Azhar and Aurelien Rodriguez and Armand Joulin and Edouard Grave and Guillaume Lample},
      year={2023},
      eprint={2302.13971},
      archivePrefix={arXiv},
      primaryClass={cs.CL},
      url={https://arxiv.org/abs/2302.13971}, 
}

@misc{jiang2024mixtralexperts,
      title={Mixtral of Experts}, 
      author={Albert Q. Jiang and Alexandre Sablayrolles and Antoine Roux and Arthur Mensch and Blanche Savary and Chris Bamford and Devendra Singh Chaplot and Diego de las Casas and Emma Bou Hanna and Florian Bressand and Gianna Lengyel and Guillaume Bour and Guillaume Lample and Lélio Renard Lavaud and Lucile Saulnier and Marie-Anne Lachaux and Pierre Stock and Sandeep Subramanian and Sophia Yang and Szymon Antoniak and Teven Le Scao and Théophile Gervet and Thibaut Lavril and Thomas Wang and Timothée Lacroix and William El Sayed},
      year={2024},
      eprint={2401.04088},
      archivePrefix={arXiv},
      primaryClass={cs.LG},
      url={https://arxiv.org/abs/2401.04088}, 
}

@misc{bonato2025repsrecycledentropypacket,
      title={REPS: Recycled Entropy Packet Spraying for Adaptive Load Balancing and Failure Mitigation}, 
      author={Tommaso Bonato and Abdul Kabbani and Ahmad Ghalayini and Michael Papamichael and Mohammad Dohadwala and Lukas Gianinazzi and Mikhail Khalilov and Elias Achermann and Daniele De Sensi and Torsten Hoefler},
      year={2025},
      eprint={2407.21625},
      archivePrefix={arXiv},
      primaryClass={cs.NI},
      url={https://arxiv.org/abs/2407.21625}, 
}

@misc{dalcin2018fastparallelmultidimensionalfft,
      title={Fast parallel multidimensional FFT using advanced MPI}, 
      author={Lisandro Dalcin and Mikael Mortensen and David E Keyes},
      year={2018},
      eprint={1804.09536},
      archivePrefix={arXiv},
      primaryClass={cs.DC},
      url={https://arxiv.org/abs/1804.09536}, 
}

\end{document}